\documentclass[3p,number,sort&compress]{elsarticle}

\journal{Annals of Physics}

\usepackage{amsmath}
\usepackage{amsfonts}
\usepackage{bbold,yfonts}
\usepackage{graphicx}
\usepackage{amssymb,mathrsfs}
\usepackage{upgreek}

\newcommand{\bs}[1]{\boldsymbol{#1}}

\begin{document}

\begin{frontmatter}

\title{Electronic hydrodynamics in graphene}


\author[kit,mifi]{Boris N. Narozhny}
\ead{boris.narozhny@kit.edu}

\address[kit]{Institut f\"ur Theorie der Kondensierten Materie,
  Karlsruher Institut f\"ur Technologie, 76131 Karlsruhe, Germany}
\address[mifi]{National Research Nuclear University MEPhI 
  (Moscow Engineering Physics Institute), 115409 Moscow, Russia}

\begin{abstract}
In this paper I report a pedagogical derivation of the unconventional
electronic hydrodynamics in graphene on the basis of the kinetic
theory. While formally valid in the weak coupling limit, this approach
allows one to derive the unconventional hydrodynamics in the system
which is neither Galilean- nor Lorentz-invariant, such that
hydrodynamic equations cannot be inferred from symmetry arguments. I
generalize earlier work to include external magnetic fields and give
explicit expressions for dissipative coefficients, the shear viscosity
and electrical conductivity. I also compare the resulting theory with
relativistic hydrodynamics.
\end{abstract}


\end{frontmatter}

%\linenumbers

\section*{}

Electronic hydrodynamics has evolved into a fast paced field with
multiple experimental and theoretical groups working to establish
observable effects of hydrodynamic behavior of electronic systems
across a wide range of materials
\cite{imh,imm,gal,geim4,sulp,geim3,ihn,goo,mac,kim1,geim1} (for a
comprehensive list of references see recent reviews
\cite{rev,luc}). Similarly to the usual hydrodynamics of ordinary
fluids, one can arrive at the final set of hydrodynamic equations in
several complementary ways. One way is purely phenomenological
\cite{dau6} : one writes an equation of motion for a small element of
the fluid, a continuity equation expressing conservation of ``mass''
(or ``matter''), and (at least for an ideal fluid) an adiabaticity
condition. The resulting five (in three dimensions) differential
equations determine the five macroscopic quantities characterizing the
(one-component) fluid, e.g., the velocity vector, fluid density, and
pressure. Another way of deriving the hydrodynamic equations
\cite{chai} is also phenomenological in nature, but is somewhat more
general since it relies on thermodynamics rather than on classical
mechanics of a fluid element. Finally, the hydrodynamic equations can
be derived from a ``microscopic'' Boltzmann theory \cite{dau10}. While
more ``technical'' and thus conceptually straightforward, this method
suffers from the narrow applicability of the Boltzmann equation itself
(i.e. the assumption of the constituent particles being free in
between successive collisions). However, one typically extends the
validity region of the hydrodynamic theory beyond that of the kinetic
equation by assumption of universality: all fluids with the same
symmetry properties obey the same set of the hydrodynamic equations
regardless of whether the coupling between constituent particles is
strong (e.g., in water) or weak (e.g., in a dilute gas).

As a macroscopic theory of strongly interacting systems, hydrodynamics
might appear to be extremely attractive for condensed matter theorists
routinely dealing with problems where strong correlations invalidate
simple theoretical approaches. However, with the exception of the
early work by Gurzhi \cite{gurzhi}, not so much attention was given to
this theory until recently. The reason for this is fairly simple:
unlike water molecules, electrons in solids exist in the environment
created by a crystal lattice and hence their momentum is not
conserved. As a result, electron motion is typically diffusive, unless
the sample size is smaller than the mean free path in which case the
system is ballistic.

For most typical scattering mechanisms in solids the mean free path is
strongly temperature dependent. At low temperatures, the electrons
scatter mostly on lattice imperfections (or ``disorder'') leading to,
e.g., the residual resistance in metals. At high temperatures the main
scattering mechanism is the electron-phonon interaction. In many
conventional (or ``simple'') metals at least one of these two
scattering mechanisms is more effective than electron-electron
interaction. In terms of the associated length scales, this statement
can be formulated as ${\ell_{ee}\gg\ell_{\rm{dis}},\ell_{\rm e-ph}}$
(with the self-evident notation). On the other hand, if a material
would exist where the opposite condition were satisfied at least in
some non-negligible temperature range, then one could be justified in
neglecting the momentum non-conserving processes and applying the
hydrodynamic theory. For a long time such a material was not known
and as a result most people working in condensed matter physics
were not interested in hydrodynamics. In recent years, the situation
has drastically changed as several extremely pure materials became
available bringing electronic hydrodynamics within experimental
reach. The best known such material is graphene
\cite{rev,luc,kats,geim3,imh,imm,gal,geim4,sulp,kim1}.

The purpose of this paper is to provide a pedagogical derivation of
the unconventional hydrodynamics in graphene. Low energy excitations
in graphene have linear dispersion and hence are not
Galilean-invariant. Their motion is restricted to the two-dimensional
(2D) plane of the graphene layer, but they are coupled via the
classical, three-dimensional (3D) Coulomb's interaction, such that the
electronic system in graphene is not Lorentz-invariant either. As a
result, one cannot simply apply either the usual or relativistic
hydrodynamics. Instead, one has to derive the hydrodynamics equations
for the electronic system in graphene from scratch
\cite{shsch,kash,mfss,alf,msf,hydro0,hydro1,julia,julia1,ks19}. The
resulting set of equations forms the ``unconventional'' hydrodynamics
in graphene.

\section{Kinetic theory of Dirac fermions in graphene}

For the purposes of this paper, I will assume the existence of a
parameter range where the low energy excitations in graphene can be
described by a kinetic (Boltzmann) equation (see Ref. \cite{kam} for a
detailed discussion and derivation from the quantum many-body theory).
I will further assume that at least some part of this parameter range
overlaps with the applicability region of the hydrodynamic theory. In
that region, the hydrodynamic equations can be derived from the
kinetic theory. The extension of the resulting theory beyond the
applicability region of the kinetic equation can then be justified by
the assumption of universality of the hydrodynamic approach.

While the kinetic theory is often used to describe electronic
transport in doped graphene \cite{can,das}, it was shown to be
inapplicable only in a region around the neutrality point \cite{aka,mish}
that is exponentially small in the dimensionless conductivity (in
units $e^2/h$) of graphene. Recent measurements \cite{gal} show this
quantity to be of order $10$ at the ``hydrodynamic'' temperatures,
$T>100$K, justifying the use of the kinetic approach to neutral
graphene at such (relatively high) temperatures.

The necessary condition for the validity of the hydrodynamics is that
the electron-electron interaction is the dominant scattering mechanism
in the system such that the typical length scale corresponding to
electron-electron interaction, $\ell_{ee}$, is the shortest length in
the problem
\begin{equation}
\label{tc}
\ell_{ee} \ll \ell_{\rm dis}, \ell_{\rm e-ph}, \ell_R , \,\, {\rm etc}.
\end{equation}
Here $\ell_{\rm dis}$, $\ell_{\rm e-ph}$, and $\ell_R$ are the length
scales characterizing disorder scattering, electron-phonon
interaction, and quasiparticle recombination \cite{alf} processes. All
other scattering mechanisms are encoded in ``etc''. Lowering
temperature towards zero, $\ell_{ee}$ is expected to diverge, while
$\ell_{\rm dis}$ is not. As a result, the inequality (\ref{tc}) can be
expected to be fulfilled at temperatures which are high enough to
justify the use of the kinetic equation even at charge neutrality (but
not too high, so that the electron-phonon interaction could still be
considered as subleading).

The general form of the kinetic equation \cite{dau10} can be seen as a
formal equality between the Liouville's operator and the collision
integral. In a two-band electronic system, the kinetic equation can be
written as
\begin{subequations}
\label{be}
\begin{equation}
\label{ke}
{\cal L}f = {\rm St}_{ee} [f] + {\rm St}_{R} [f] - \frac{f-\langle f \rangle_\varphi}{\tau_{\rm dis}},
\end{equation}
with (here $\bs{E}$ and $\bs{B}$ are the electric and magnetic fields)
\begin{equation}
\label{l}
{\cal L} = \partial_t + \bs{v}\!\cdot\!\bs{\nabla}_{\bs{r}} + 
(e\bs{E} + \frac{e}{c} \bs{v}\!\times\!\bs{B})\!\cdot\!\bs{\nabla}_{\bs{k}}.
\end{equation}
\end{subequations}
Labeling single-particle states by the band index ${\lambda=\pm}$ and
the momentum $\bs{k}$, one can denote the distribution function by
${f=f_{\lambda\bs{k}}}$. The collision integral comprises three parts:
${\rm{St}}_{ee}[f]$ describes electron-electron interaction,
${\rm{St}}_{R}[f]$ -- electron-hole recombination, while the remaining
term in Eq.~(\ref{ke}) describes disorder scattering. The latter
involves angular averaging defined as
\begin{equation}
\label{anav}
\langle f \rangle_\varphi = \int\limits_{-\pi}^\pi \frac{d\varphi}{2\pi}
f_{\lambda\bs{k}},
\end{equation}
where $\varphi$ is the polar angle describing the direction of $\bs{k}$.

The form of the Liouville's operator is independent of whether the
underlying microscopic physics is classical or quantum, at least as
long as there is no spin-orbit interaction. The collision integral is
more sensitive to the microscopic details of the system. In
particular, the $\tau$-approximation employed in Eq.~(\ref{ke}) to
describe disorder scattering is almost certainly an
oversimplification. Even then, $\tau_{\rm dis}$ is a model-dependent
function of energy \cite{kats}. However in the limit of weak disorder,
required by Eq.~(\ref{tc}), one may assign a particular large value to
$\tau_{\rm{dis}}$ (as determined by the temperature, $T$, and chemical
potential, $\mu$) such that most physical observables (with the
notable exception of thermal conductivity) will be insensitive to the
choice of the impurity model. Furthermore, the collision integral in
Eq.~(\ref{ke}) disregards any ``quantum'' or ``interference''
corrections to quasiparticle transport
\cite{ala,aag,zna,al1,lee,df1,df2,dfe1,dfe2}. Below, I treat
${\rm{St}}_{ee}[f]$ at the Golden Rule level. While affecting the
numerical values of theoretical estimates for dissipative
coefficients, this approximation has no bearing on the form of
hydrodynamic equations, which is the main goal of this derivation.

In this paper, I derive the hydrodynamics equations in graphene
following the standard textbook steps \cite{dau10}: (i) integrating
the kinetic equation (\ref{be}), I obtain continuity equations
expressing conservation of the particle number (or electric charge),
energy, and momentum; (ii) assuming local equilibrium, I relate the
quantities appearing in the continuity equations to macroscopic
quantities characterizing the electronic fluid (i.e., particle and
energy densities and the flow velocity) and thus determine the
equations of ideal hydrodynamics in graphene (i.e., the generalization
of the usual Euler equation); (iii) using an approximate solution to
the kinetic equation, I establish the leading dissipative corrections
to the ideal hydrodynamics and establish the generalization of the main
equation of the usual hydrodynamics, i.e., the Navier-Stokes equation.
At the latter step I determine the explicit expressions for the
dissipative coefficients, such as shear viscosity and quantum
conductivity.

\section{Ideal hydrodynamics in graphene}

\subsection{Continuity equations}

Hydrodynamics is a direct consequence of conservation laws. These are
commonly expressed in terms of continuity equations, which can be
written on a phenomenological basis \cite{dau6,chai}. Integrating the
kinetic equation, one can not only ``derive'' the continuity
equations, but also give explicit expressions for the corresponding
densities and currents in terms of the distribution function.

\subsubsection{Particle number conservation}

The equation most commonly known as ``the continuity equation''
\cite{dau6} expresses conservation of ``mass'' or ``matter'': the
amount of fluid in any given volume can be changed only by means of
fluid flow through the volume boundary. In an electronic system, this
is equivalent to conservation of electric charge. Within the kinetic
theory, this conservation law is manifested in the vanishing
of the integrated collision integral in (or the right-hand side of)
the kinetic equation
\[
N\sum_\lambda\int\frac{d^2k}{(2\pi)^2} \left[
{\rm St}_{ee} [f] + {\rm St}_{R} [f] - \frac{f-\langle f \rangle_\varphi}{\tau_{\rm dis}}
\right]=0.
\]
Here $\lambda=\pm$ is the band index, $\bs{k}$ is the momentum
labeling single-particle states, and $N$ is the degeneracy factor (in
real graphene $N=4$ due to spin and valley degeneracy).

The continuity equation can be obtained by integrating the kinetic
equation (\ref{ke}) and has the usual form \cite{dau6,dau10}. The only
subtle point arising in two-band systems is the treatment of the
formally infinite number of particles in the filled band. However,
assuming the contribution of the filled band to be constant, one can
immediately see that it vanishes upon differentiation and does
not contribute to the continuity equation.

Consider first the time-derivative term in Eq.~(\ref{l}). Integrating
this term over all states yields
\[
N\!\sum_\lambda\!\int\!\frac{d^2k}{(2\pi)^2} \partial_t f_{\lambda\bs{k}} =
\partial_t \, N \!\!\int\!\!\frac{d^2k}{(2\pi)^2} \left[
f_{+,\bs{k}} \!-\! \left(1\!-\!f_{-,\bs{k}}\right)\right] 
=
\partial_t(n_+\!-\!n_-) = \partial_t n,
\]
defining the charge density, $n$, (up to the factor of electric
charge). The definitions of the numbers of charge carriers in the two bands,
$n_\pm$, are given in \ref{qds}.

Similarly, the gradient term can be integrated as
\[
N\!\sum_\lambda\!\int\!\frac{d^2k}{(2\pi)^2} \bs{v}\!\cdot\!\bs{\nabla}_{\bs{r}} f_{\lambda\bs{k}} 
\!=\!
\partial_i \, N \!\int\!\!\frac{d^2k}{(2\pi)^2}\! \left[
v^i_{+,\bs{k}} f_{+,\bs{k}}\!-\!v^i_{-,\bs{k}} \left(1\!-\!f_{-,\bs{k}}\right)
\right] 
=
\partial_i\left(j^i_+\!-\!j^i_-\right)
=
\bs{\nabla}_{\bs{r}}\!\cdot\!\bs{j},
\]
defining the electric current, $\bs{j}$ (up to the factor of the
electron charge; see \ref{qpcs} for explicit definitions of the
quasiparticle currents, $\bs{j}_\pm$ and $\bs{j}$).

Charge conservation requires that the electric field does not affect
the continuity relation. Technically, this is expressed by means of
the vanishing integral
\[
e\bs{E}\!\cdot N\!\sum_\lambda\!\int\!\frac{d^2k}{(2\pi)^2} \bs{\nabla}_{\bs{k}} f_{\lambda\bs{k}} = 
e\bs{E}\!\cdot N\!\!\int\!\frac{d^2k}{(2\pi)^2}
\left[ \bs{\nabla}_{\bs{k}} f_{+\bs{k}}- \bs{\nabla}_{\bs{k}} \left(1\!-\!f_{-\bs{k}}\right)\right] =
0.
\]
The situation with the magnetic field is more involved. Integrating
the Lorentz term in Eq.~(\ref{ke}), one finds
\[
\epsilon^{\alpha\beta\gamma} B^\gamma
N\!\sum_\lambda\!\int\!\frac{d^2k}{(2\pi)^2} v^\beta_{\lambda\bs{k}} 
\frac{\partial f_{\lambda\bs{k}}}{\partial k^\alpha}
=
- \epsilon^{\alpha\beta\gamma} B^\gamma
N\!\sum_\lambda\!\int\!\frac{d^2k}{(2\pi)^2} f_{\lambda\bs{k}} 
\frac{\partial v^\beta_{\lambda\bs{k}}}{\partial k^\alpha}.
\]
For any rotationally invariant spectrum, velocity and momentum have
the same direction and the latter expression vanishes
\[
\partial v^\beta_{\lambda\bs{k}}/\partial k^\alpha \propto \delta_{\alpha\beta}, \qquad
\epsilon^{\alpha\beta\gamma} \delta_{\alpha\beta} = 0.
\]
Systems with anisotropic spectra should be considered separately. Such
analysis is beyond the scope of this paper. Whatever the spectrum, the
Lorentz force cannot violate charge conservation.

Combining the above contributions, I find the standard continuity
equation (usually, the continuity equation is expressed in terms of
the charge density and electric current, which differ from the
quantities $n$ and $\bs{j}$ by a multiplicative factor of the electric
charge)
\begin{subequations}
\label{ce}
\begin{equation}
\label{cen}
\partial_t n + \bs{\nabla}_{\bs{r}}\!\cdot\!\bs{j} = 0,
\end{equation}
which is valid for any electronic system (even if the kinetic equation
itself is not).

Multiplying the kinetic equation by $\lambda$ and integrating over all
states, one can find a similar equation for the imbalance current,
$\bs{j}_I$, [see Eq.~(\ref{ji})]
\begin{equation}
\label{ceni}
\partial_t n_I + \bs{\nabla}_{\bs{r}}\!\cdot\!\bs{j}_I = - \frac{n_I\!-\!n_I^{(0)}}{\tau_R},
\end{equation}
\end{subequations}
where the right-hand side describes the recombination processes
\cite{alf} (within the $\tau$-approximation; $n_I$ is the imbalance
density, see \ref{qds}, and $n_I^{(0)}$ is the equilibrium imbalance
density). Technically this term appears from the integration of the
collision integral ${\rm St}_{R} [f]$, which does not conserve the
number of particles in each band individually. In monolayer graphene,
the dominant process contributing to quasiparticle recombination is
the impurity-assisted electron-phonon scattering \cite{srl,hydro1}.

\subsubsection{Energy conservation}

In two-band systems with unbound (from below) spectrum, one has to
define the energy density relative to the (formally infinite) energy
of the filled valence band, see \ref{neapp}.

Similarly to the particle number conservation, energy conservation
leads to the vanishing integral
\[
N\!\sum_\lambda\!\int\!\frac{d^2k}{(2\pi)^2} \; \epsilon_{\lambda\bs{k}} \left[
{\rm St}_{ee} [f] + {\rm St}_R [f]- \frac{f-\langle f \rangle_\varphi}{\tau_{\rm dis}}
\right] =0.
\]
Multiplying the kinetic equation (\ref{ke}) by the energy and
integrating over all single-particle states leads to the continuity
equation for the energy density \cite{dau6,dau10}.

Since the quasiparticle energies and the energy of the filled valence
band (\ref{neinf}) are independent of time, integrating the first term
in the Liouville's operator yields
\[
N\!\sum_\lambda\!\int\!\frac{d^2k}{(2\pi)^2} \epsilon_{\lambda\bs{k}} \partial_t f_{\lambda\bs{k}} =
\partial_t \; N \!\int\!\!\frac{d^2k}{(2\pi)^2} \epsilon_{\lambda\bs{k}} f_{\lambda\bs{k}}
= \partial_t n_E,
\]
where the energy density, $n_E$, is defined in Eq.~(\ref{ne}).

The integrated gradient term in the Liouville's operator defines the
energy current, $\bs{j}_E$ [see also Eq.~(\ref{je})]
\[
N\!\sum_\lambda\!\int\!\!\frac{d^2k}{(2\pi)^2} 
\epsilon_{\lambda\bs{k}} \bs{v}_{\lambda\bs{k}}\!\cdot\!\bs{\nabla}_{\bs{r}} f_{\lambda\bs{k}} =
\bs{\nabla}_{\bs{r}}\!\cdot\! N \!\int\!\!\frac{d^2k}{(2\pi)^2}\! 
\epsilon_{\lambda\bs{k}} \bs{v}_{\lambda\bs{k}} f_{\lambda\bs{k}}
= \bs{\nabla}_{\bs{r}}\!\cdot\!\bs{j}_E.
\]

The electric field acting on an electronic system leads to Joule's
heating. Integrating the electric field term in the kinetic equation
one finds
\[
e\bs{E} \cdot N\!\sum_\lambda\!\int\!\!\frac{d^2k}{(2\pi)^2} \epsilon_{\lambda\bs{k}}
\bs{\nabla}_{\bs{k}} f_{\lambda\bs{k}}
= - e\bs{E} \cdot N\!\sum_\lambda\!\int\!\!\frac{d^2k}{(2\pi)^2} \bs{v}_{\lambda\bs{k}}
f_{\lambda\bs{k}} 
= - e \bs{E}\cdot\bs{j}.
\]

Finally, the Lorentz force cannot lead to any change of energy since
it does not do any work. Indeed, integrating the Lorentz term in Eq.~(\ref{l}), one
finds (for a rotationally invariant system)
\[
\epsilon^{\alpha\beta\gamma} B^\gamma
N\!\sum_\lambda\!\int\!\frac{d^2k}{(2\pi)^2} \epsilon_{\lambda\bs{k}} v^\beta_{\lambda\bs{k}} 
\frac{\partial f_{\lambda\bs{k}}}{\partial k^\alpha} 
=
- \epsilon^{\alpha\beta\gamma} B^\gamma
N\!\sum_\lambda\!\int\!\frac{d^2k}{(2\pi)^2} f_{\lambda\bs{k}}\epsilon_{\lambda\bs{k}} 
\frac{\partial v^\beta_{\lambda\bs{k}}}{\partial k^\alpha}.
\]
Here the energy $\epsilon_{\lambda\bs{k}}$ is not differentiated since
this would just yield a velocity and a cross product of two velocities
is zero. The last term is similar to that in the continuity equation.

Combining all of the above, I find the continuity equation for the
energy density
\begin{equation}
\label{cene}
\partial_t n_E + \bs{\nabla}_{\bs{r}}\!\cdot\!\bs{j}_E = e \bs{E}\cdot\bs{j},
\end{equation}
where the right-hand side describes the Joule's heat.

\subsubsection{Momentum conservation}

Multiplying the kinetic equation by the momentum and integrating over
all states, one finds that the collision integrals due to
electron-electron interaction and quasiparticle recombination vanish
\[
N\!\sum_\lambda\!\int\!\frac{d^2k}{(2\pi)^2} \; \bs{k} \;
\left({\rm St}_{ee} [f] + {\rm St}_R [f]\right) =0.
\]
Impurity scattering, however, may relax momentum so that (unlike in
the two previous cases) the impurity contribution to the collision
integral does not vanish. Within the $\tau$-approximation [see
  Eq.~(\ref{ke})], one finds
\[
N\!\sum_\lambda\!\int\!\!\frac{d^2k}{(2\pi)^2} \bs{k}
\frac{f\!-\!\langle f \rangle_\varphi}{\tau_{\rm dis}}
=
\frac{1}{\tau_{\rm dis}}
N\!\sum_\lambda\!\int\!\frac{d^2k}{(2\pi)^2} \bs{k} f_{\lambda\bs{k}} 
=
\frac{\bs{n}_{\bs{k}}}{\tau_{\rm dis}},
\]
where the momentum density, $\bs{n}_{\bs{k}}$, is defined in \ref{jeapp}.

The time derivative term in Eq.~(\ref{l}) is treated same as before
\[
N\!\sum_\lambda\!\int\!\!\frac{d^2k}{(2\pi)^2} \bs{k} \partial_t f_{\lambda\bs{k}}
=
\partial_t N\!\sum_\lambda\!\int\!\!\frac{d^2k}{(2\pi)^2} \bs{k} f_{\lambda\bs{k}}
= \partial_t \bs{n}_{\bs{k}}.
\]

Integrating the gradient term, one finds the momentum flux tensor (\ref{pab})
\[
N\!\sum_\lambda\!\int\!\!\frac{d^2k}{(2\pi)^2} 
k^\alpha v^\beta_{\lambda\bs{k}}\nabla^\beta_{\bs{r}} f_{\lambda\bs{k}} 
=
\nabla^\beta_{\bs{r}}
N\!\sum_\lambda\!\int\!\!\frac{d^2k}{(2\pi)^2} 
k^\alpha v^\beta_{\lambda\bs{k}} f_{\lambda\bs{k}}
=
\nabla^\beta_{\bs{r}} \Pi_E^{\alpha\beta}.
\]

The external forces can also change the momentum density. The
electric field term can be integrated as follows
\begin{eqnarray*}
&&
eE^\beta N\!\sum_\lambda\!\int\!\!\frac{d^2k}{(2\pi)^2}  k^\alpha 
\nabla^\beta_{\bs{k}} f_{\lambda\bs{k}} 
=
eE^\beta N\!\!\int\!\!\frac{d^2k}{(2\pi)^2}  k^\alpha \nabla^\beta_{\bs{k}}
\left[ f_{+,\bs{k}} \!-\! (1\!-\!f_{-,\bs{k}})\right]
\\
&&
\\
&&
\qquad\qquad\qquad\qquad\qquad\qquad
=
- eE^\beta N\!\!\int\!\!\frac{d^2k}{(2\pi)^2} \left[ f_{+,\bs{k}} \!-\! (1\!-\!f_{-,\bs{k}})\right]
\nabla^\beta_{\bs{k}} k^\alpha
=
- e n E^\alpha.
\end{eqnarray*}
The Lorentz force can change the direction of momentum. Integrating
the Lorentz term similarly the above I find
\[
\epsilon^{\alpha\beta\gamma} B^\gamma
N\!\sum_\lambda\!\int\!\!\frac{d^2k}{(2\pi)^2} k^\mu v^\beta_{\lambda\bs{k}} 
\frac{\partial f_{\lambda\bs{k}}}{\partial k^\alpha} 
=
- \epsilon^{\alpha\beta\gamma} B^\gamma
N\!\sum_\lambda\!\int\!\!\frac{d^2k}{(2\pi)^2} f_{\lambda\bs{k}} \!
\left[ \delta_{\mu\alpha}v^\beta_{\lambda\bs{k}} 
  \!+\! k^\mu \frac{\partial v^\beta_{\lambda\bs{k}}}{\partial k^\alpha} \right]
=
-\bs{j}\!\times\!\bs{B},
\]
where at the last step I relied on rotational invariance.

Finally, I find the following equation
\begin{equation}
\label{cek}
\partial_t n^\alpha_{\bs{k}} + \nabla^\beta_{\bs{r}} \Pi_E^{\alpha\beta}
- e n E^\alpha - \frac{e}{c} \left[\bs{j}\!\times\!\bs{B}\right]^\alpha =
- \frac{n^\alpha_{\bs{k}}}{\tau_{\rm dis}},
\end{equation}
which differs from the usual continuity equation for the momentum
density \cite{dau10} by the presence of the electromagnetic fields and
the weak disorder scattering term. The latter has to be small [see
  Eq.~(\ref{tc})], otherwise the discussion of hydrodynamics makes no
sense. However, the arguments leading to Eq.~(\ref{cek}) are rather
general: they do not rely on Eq.~(\ref{tc}) and are justified in the
whole applicability region of the kinetic equation (\ref{be}).

\subsubsection{Summary of the continuity equations}

To summarize this section, I list all four continuity equations for a
two-band electronic system (the above arguments are not specific to
graphene)
\begin{subequations}
\label{ces}
\begin{equation}
\label{cen1}
\partial_t n + \bs{\nabla}_{\bs{r}}\!\cdot\!\bs{j} = 0,
\end{equation}
\begin{equation}
\label{ceni1}
\partial_t n_I + \bs{\nabla}_{\bs{r}}\!\cdot\!\bs{j}_I = - \frac{n_I\!-\!n_I^{(0)}}{\tau_R},
\end{equation}
\begin{equation}
\label{cene1}
\partial_t n_E + \bs{\nabla}_{\bs{r}}\!\cdot\!\bs{j}_E = e \bs{E}\cdot\bs{j},
\end{equation}
\begin{equation}
\label{cek1}
\partial_t n^\alpha_{\bs{k}} + \nabla^\beta_{\bs{r}} \Pi_E^{\alpha\beta}
- e n E^\alpha - \frac{e}{c} \left[\bs{j}\!\times\!\bs{B}\right]^\alpha =
- \frac{n^\alpha_{\bs{k}}}{\tau_{\rm dis}}.
\end{equation}
\end{subequations}

\subsection{Local equilibrium}

The main underlying assumption of the hydrodynamic theory is that of
the local equilibrium established by the electron-electron collisions
\cite{dau10} on length scales much larger than $\ell_{ee}$. The
corresponding distribution function can be written as \cite{hydro1}
\begin{equation}
\label{le}
f^{(0)}_{\lambda\bs{k}} (\bs{r}) =
\left\{
1+\exp\left[\frac{\epsilon_{\lambda\bs{k}}-\mu_\lambda(\bs{r}) - 
\bs{u}(\bs{r})\!\cdot\!\bs{k}}{T(\bs{r})}\right]
\right\}^{-1},
\end{equation}
where ${\mu_\lambda(\bs{r})}$ is the local chemical potential and
$\bs{u}(\bs{r})$ is the hydrodynamic (or ``drift'') velocity.

In order to use the distribution function (\ref{le}) in practical
calculations, one needs to specify the quasiparticle spectrum (the
above general form of the continuity equations is valid for an
arbitrary two-band electronic system). In this paper, I will employ
the following notations for the Dirac spectrum (the chirality
$\lambda=\pm$ distinguishes the conduction and valence bands)
\begin{equation}
\label{eg}
\epsilon_{\lambda\bs{k}} = \lambda v_g k,
\end{equation}
and velocities (hereafter $\bs{e}_{\bs{a}}$ denotes a unit vector in
the direction $\bs{a}$)
\begin{equation}
\label{vg}
\bs{v}_{\lambda\bs{k}}=\lambda v_g \frac{\bs{k}}{k}, \quad
\bs{e}_{\bs{v}} = \bs{e}_{\bs{k}},
\quad
\bs{k} = \frac{\lambda k}{v_g}\bs{v}_{\lambda\bs{k}}
=\frac{\epsilon_{\lambda\bs{k}}\bs{v}_{\lambda\bs{k}}}{v_g^2}.
\end{equation}

Using the distribution function (\ref{le}) and the Dirac spectrum
(\ref{eg}) in the definitions (see \ref{appA}) of the hydrodynamic
quantities featuring in the continuity equations (\ref{ces}), one can
arrive at the equilibrium expressions for the macroscopic densities
and currents. In particular, the quasiparticle currents (\ref{j}) and
(\ref{ji}) can be expressed in terms of $\bs{u}$ and the corresponding
densities, as one might expect
\begin{equation}
\label{hj0s}
\bs{j} = n \bs{u},
\qquad
\bs{j}_{I} = n_{I} \bs{u},
\end{equation}
while the ``charge'' and ``imbalance'' densities defined in Eqs.~(\ref{den})
are given by
\begin{subequations}
\label{n0s}
\begin{equation}
\label{n0}
n = n_{+}\! -\! n_{-} = \frac{NT^2}{2\pi v_g^2}
\frac{\tilde n}{\left(1\!-\!u^2/v_g^2\right)^{3/2}},
\qquad
\tilde n = - {\rm Li}_2\left(-e^{\mu_+/T}\right) + {\rm Li}_2\left(-e^{\mu_-/T}\right),
\end{equation}
\begin{equation}
\label{nI0_0}
n_{I} = n_{+} \!+\! n_{-} = \frac{NT^2}{2\pi v_g^2}
\frac{\tilde n_I}{\left(1\!-\!u^2/v_g^2\right)^{3/2}},
\qquad
\tilde n_I = - {\rm Li}_2\left(-e^{\mu_+/T}\right) - {\rm Li}_2\left(-e^{\mu_-/T}\right),
\end{equation}
where ${\rm Li}_n(z)$ is the polylogarithm. For the simplest case $\mu_\pm=\mu$, the
total quasiparticle density (\ref{nI0_0}) simplifies
\begin{equation}
\label{tnI0}
\tilde n_I(x=\mu/T) = \frac{x^2}{2}+\frac{\pi^2}{6}.
\end{equation}
\end{subequations}

The energy current (\ref{je}) can be expressed in terms of
the energy density (\ref{ne})
\begin{equation}
\label{hje}
\bs{j}_{E} = \frac{3n_{E}\bs{u}}{2\!+\!u^2/v_g^2},
\end{equation}
where the energy density is given by
\begin{equation}
\label{ne0}
n_{E} = \frac{NT^3}{\pi v_g^2}
\frac{1\!+\!u^2/(2v_g^2)}{\left(1\!-\!u^2/v_g^2\right)^{5/2}} \tilde{n}_E,
\qquad
\tilde{n}_E = - {\rm Li}_3\left(-e^{\mu_+/T}\right) - {\rm Li}_3\left(-e^{\mu_-/T}\right).
\end{equation}

The momentum density $\bs{n}_{\bs{k}}$ is proportional to the energy
current, see Eq.~(\ref{nk}). The momentum flux tensor (\ref{pab}) is
also expressed in terms of the energy density (\ref{ne0})
\begin{equation}
\label{hpi}
\Pi_{E}^{\alpha\beta} 
= 
\frac{n_{E}}{2\!+\!u^2/v_g^2}
\left[\delta_{\alpha\beta} \left(1\!-\!\frac{u^2}{v_g^2}\right) 
\!+\! \frac{3u_\alpha u_\beta}{v_g^2}\right].
\end{equation}

Finally, the local equilibrium distribution function (\ref{le}) can be
used to compute the thermodynamic quantities. Introducing the linear
combinations of the two chemical potentials $\mu_\pm$ (i.e. defining
the thermodynamic variables conjugated to $n$ and $n_I$),
\begin{equation}
\label{mu}
\mu=\frac{\mu_++\mu_-}{2}, \quad
\mu_I=\frac{\mu_+-\mu_-}{2}
\qquad\Leftrightarrow\qquad
\mu_+=\mu+\mu_I, \quad
\mu_-=\mu-\mu_I,
\end{equation}
I can define the compressibilities
\begin{subequations}
\label{dndmu}
\begin{equation}
\label{dos0}
\frac{\partial n}{\partial \mu}
=
\frac{NT}{2\pi v_g^2}\left[
  \ln\left(1\!+\!e^{\mu_+/T}\right)\!+\!\ln\left(1\!+\!e^{-\mu_-/T}\right)
  \right]\!,
\end{equation}
\begin{equation}
\label{dosI0}
\frac{\partial n_{I}}{\partial \mu}
\!=\!
\frac{NT}{2\pi v_g^2}\!\left[
  \ln\left(1\!+\!e^{\mu_+/T}\right)\!-\!\ln\left(1\!+\!e^{-\mu_-/T}\right)\!
  \right]\!.
\end{equation}
In the simplest case ${\mu_\pm=\mu}$ (i.e., ${\mu_I=0}$) the expression
(\ref{dosI0}) simplifies to
\begin{equation}
\label{dosI0mu}
\left.\frac{\partial n_{I}}{\partial \mu}\right|_{\mu_\pm=\mu}
\!=\!
\frac{N\mu}{2\pi v_g^2},
\end{equation}
in obvious agreement with Eq.~(\ref{tnI0}). In the same case, the
compressibility (\ref{dos0}) can be re-written with the help of the
short-hand notation
\begin{equation}
\label{dost}
\frac{\partial n}{\partial \mu}
=
\frac{N{\cal T}}{2\pi v_g^2},
\end{equation}
with
\begin{equation}
\label{t}
{\cal T} = T
\left[\ln\left(1\!+\!e^{\mu/T}\right)+\ln\left(1\!+\!e^{-\mu/T}\right)\right]
= 2T\ln\left[2\cosh\frac{\mu}{2T}\right].
\end{equation}
\end{subequations}
The hydrodynamic pressure (\ref{p}) is proportional to the energy density
\begin{equation}
\label{pg}
P = n_{E} \frac{1\!-\!u^2/v_g^2}{2\!+\!u^2/v_g^2},
\end{equation}
and hence for the enthalpy one finds
\begin{equation}
\label{enth}
W=n_{E}+P = \frac{3 n_{E}}{2\!+\!u^2/v_g^2}.
\end{equation}
As a result, the expressions for the energy current (\ref{hje}) and
momentum flux tensor (\ref{hpi}) can be re-written as
\begin{equation}
\label{hjepiw}
\bs{j}_{E} = W\bs{u},
\qquad
\Pi_{E}^{\alpha\beta} 
= 
P\delta_{\alpha\beta} + \frac{W}{v_g^2}u_\alpha u_\beta.
\end{equation}

\subsection{Generalized Euler equation}

Substituting the above hydrodynamic quantities into the continuity
equations (\ref{ces}), I obtain the equations of the ideal
hydrodynamics in graphene. Consider first the equation (\ref{cek1})
representing momentum conservation. Using Eqs.~(\ref{hjepiw}) and
(\ref{nk}), I find for the two derivative terms in Eq.~(\ref{cek1}):
\[
\partial_t \bs{n}_{\bs{k}} = v_g^{-2} \partial_t (W\bs{u}) =
v_g^{-2} (W \partial_t \bs{u} + \bs{u} \partial_t W);
\]
\[
\nabla_{\bs{r}}^\beta\Pi_E^{\alpha\beta} 
= 
\nabla_{\bs{r}}^\beta (P\delta_{\alpha\beta}+v_g^{-2}Wu^\alpha u^\beta)
=
\nabla_{\bs{r}}^\alpha P \!+\! v_g^{-2} \left[W (\bs{u}\!\cdot\!\bs{\nabla}) u^\alpha 
\!+\! u^\alpha \bs{\nabla}\!\cdot\!(W\bs{u})\right].
\]
The last term can be found from Eq.~(\ref{cene1}) where one can use 
the energy density from Eq.~(\ref{enth}) and the energy current from
Eq.~(\ref{hje})
\[
\partial_t(W\!-\!P) + \bs{\nabla}\!\cdot\!(W\bs{u}) = e\bs{E}\!\cdot\!\bs{j}.
\]
Combining the above three equalities, I find 
\[
\partial_t n_{\bs{k}}^\alpha + \nabla_{\bs{r}}^\beta\Pi_E^{\alpha\beta}
=
v_g^{-2} W(\partial_t\!+\!\bs{u}\!\cdot\!\bs{\nabla})u^\alpha \!+\! \nabla_{\bs{r}}^\alpha P
\!+\!
v_g^{-2} u^\alpha\left[ \partial_t P \!+\! e\bs{E}\!\cdot\!\bs{j}\right].
\]
As a result, I find the generalization of the Euler equation for
graphene taking into account weak disorder and external
electromagnetic fields (cf. the standard Euler equation \cite{dau6}
comprising the first three terms on the left-hand side)
\begin{eqnarray}
\label{eq0g}
W(\partial_t+\bs{u}\!\cdot\!\bs{\nabla})\bs{u}
+
v_g^2 \bs{\nabla} P
+
\bs{u} \partial_t P 
+
e(\bs{E}\!\cdot\!\bs{j})\bs{u} 
=
v_g^2 en\bs{E}
+
v_g^2 \frac{e}{c} \bs{j}\!\times\!\bs{B}
-
\frac{W\bs{u}}{\tau_{{\rm dis}}}.
\end{eqnarray}

Combined with the continuity equations (\ref{ces}), Eq.~(\ref{eq0g})
describes the ideal electron-hole fluid in graphene. Hydrodynamics of
an ideal fluid is well-studied \cite{dau6,dau10}. In what follows, I
relate Eq.~(\ref{eq0g}) to the standard Euler equation and its
relativistic counterpart as well as consider the simplest solutions of
the ideal hydrodynamics.

\subsection{Entropy flow}
\label{egr}

Within the usual hydrodynamics, the ideal flow is isentropic
\cite{dau6} and hence one can derive a ``continuity equation'' for
entropy, which provides a definition of the entropy current. 
The entropy density of a system of fermions is
defined in terms of the distribution function as
\begin{equation}
\label{sdk}
s = - N\!\sum_\lambda\!\int\!\frac{d^2k}{(2\pi)^2} 
\left[f_{\lambda\bs{k}} \ln f_{\lambda\bs{k}}
+ (1\!-\!f_{\lambda\bs{k}})\ln(1\!-\!f_{\lambda\bs{k}})\right].
\end{equation}
Treating this integral as 
\[
s = N\!\sum_\lambda\!\int\!\frac{d^2k}{(2\pi)^2} {\cal S}[f_{\lambda\bs{k}}],
\qquad
{\cal S}[f_{\lambda\bs{k}}] = - \left[f_{\lambda\bs{k}} \ln f_{\lambda\bs{k}}
+ (1\!-\!f_{\lambda\bs{k}})\ln(1\!-\!f_{\lambda\bs{k}})\right],
\]
I can represent any derivative of $s$ in the form
\[
\frac{\partial s}{\partial z} = 
N\!\sum_\lambda\!\int\!\frac{d^2k}{(2\pi)^2} 
\frac{\partial {\cal S}[f_{\lambda\bs{k}}]}{\partial f_{\lambda\bs{k}}} 
\frac{\partial f_{\lambda\bs{k}}}{\partial z}.
\]
Multiplying the kinetic equation in the absence of external
fields by the derivative $\partial {\cal
  S}[f_{\lambda\bs{k}}]/\partial f_{\lambda\bs{k}}$ and summing over
all states I obtain a macroscopic equation
\begin{equation}
\label{ceen}
\frac{\partial s}{\partial t} + \bs{\nabla}_{\bs{r}}\!\cdot\!\bs{j}_S 
=
{\cal I}.
\end{equation}

On the left-hand side of Eq.~(\ref{ceen}), $s$ is the entropy density
(\ref{eng}) and the vector $\bs{j}_S$ can be interpreted as the
entropy current
\begin{equation}
\label{jen0}
\bs{j}_S =  N\!\sum_\lambda\!\int\!\frac{d^2k}{(2\pi)^2} \bs{v}_{\lambda\bs{k}}
{\cal S}[f_{\lambda\bs{k}}].
\end{equation}
Since the function $\cal S$ vanishes as $k\rightarrow\infty$, one may
integrate by parts:
\[
\bs{j}_S =  - N\!\sum_\lambda\!\int\!\frac{d^2k}{(2\pi)^2} \epsilon_{\lambda\bs{k}}
\bs{\nabla}_{\bs{k}} {\cal S}[f_{\lambda\bs{k}}]
=
- N\!\sum_\lambda\!\int\!\frac{d^2k}{(2\pi)^2} \epsilon_{\lambda\bs{k}}
\frac{\partial {\cal S}[f_{\lambda\bs{k}}]}{\partial f_{\lambda\bs{k}}}
\bs{\nabla}_{\bs{k}} f_{\lambda\bs{k}}.
\]
Using the explicit form of the derivative, this expression can be
re-written as
\[
\bs{j}_S =  - \frac{N}{T}\!\sum_\lambda\!\int\!\frac{d^2k}{(2\pi)^2} 
\epsilon_{\lambda\bs{k}}
\left[ \epsilon_{\lambda\bs{k}}\!-\!\mu_\lambda \!-\! \bs{u}\!\cdot\!\bs{k} \right]
\bs{\nabla}_{\bs{k}} f_{\lambda\bs{k}}
=
\frac{N}{T}\!\sum_\lambda\!\int\!\frac{d^2k}{(2\pi)^2} 
\left[
2 \epsilon_{\lambda\bs{k}} \bs{v}_{\lambda\bs{k}} 
\!-\! \mu_\lambda \bs{v}_{\lambda\bs{k}}
\!-\! (\bs{u}\!\cdot\!\bs{k}) \bs{v}_{\lambda\bs{k}}
\!-\! \bs{u} \epsilon_{\lambda\bs{k}}
\right] f_{\lambda\bs{k}},
\]
which yields upon the substitution of the definitions (\ref{js}),
(\ref{ne}), (\ref{je1}), and (\ref{pab})
\begin{equation}
\label{jenk}
j^\alpha_S = \frac{1}{T} \left[ 2 j^\alpha_E \!-\! \mu_+ j^\alpha_+ \!+\! \mu_-j^\alpha_-
\!-\! u^\beta \Pi_E^{\alpha\beta}
\!-\! u^\alpha n_E
\right].
\end{equation}
Finally, using the explicit expressions for the macroscopic
quantities in Eq.~(\ref{jenk}), I arrive at the result
\begin{equation}
\label{jen}
\bs{j}_S = s \bs{u},
\end{equation}
where $s$ is the entropy density (\ref{eng}).

Now, the right-hand side of Eq.~(\ref{ceen}) is the integrated collision integral
\[
{\cal I} = N\!\sum_\lambda\!\int\!\frac{d^2k}{(2\pi)^2} 
\frac{\partial {\cal S}[f_{\lambda\bs{k}}]}{\partial f_{\lambda\bs{k}}} 
\left[{\rm St}_{ee} [f] + {\rm St}_{R} [f] 
- \frac{f-\langle f \rangle_\varphi}{\tau_{\rm dis}}\right].
\]
Evaluating the derivative of ${\cal S}$ for the local equilibrium
distribution function (\ref{le}) explicitly, I find
\[
\frac{\partial {\cal S}[f_{\lambda\bs{k}}]}{\partial f_{\lambda\bs{k}}}
=
- \ln \frac{f_{\lambda\bs{k}}}{1\!-\!f_{\lambda\bs{k}}}
=
\ln \left[\frac{1}{f_{\lambda\bs{k}}}\!-\!1\right]=
\frac{\epsilon_{\lambda\bs{k}}\!-\!\mu_\lambda \!-\! 
\bs{u}\!\cdot\!\bs{k}}{T}.
\]
The first term does not contribute to the integral due to energy
conservation. In the simplest case, ${\mu_\pm=\mu}$, the second term
also vanishes due to charge conservation. However, if the
quasiparticle imbalance becomes important, i.e., in the presence of a
temperature gradient \cite{alf}, the second term yields a
non-vanishing contribution due to electron-hole recombination (since
the collision integral ${\rm St}_{R}[f]$ does not conserve the number
of particles in each individual band. This situation is outside of the
scope of this paper and will be considered elsewhere. Finally, the
last term yields the following contribution due to disorder scattering
(which does not conserve momentum)
\[
{\cal I} = \frac{1}{T}\bs{u}\!\cdot\!N\!\sum_\lambda\!\int\!\frac{d^2k}{(2\pi)^2}
\bs{k} \frac{f\!-\!\langle f \rangle_\varphi}{\tau_{\rm dis}}
=
\frac{\bs{u}\!\cdot\!\bs{n}_{\bs{k}}}{T\tau_{\rm dis}}.
\]

Combining the above arguments, I obtain the final form of the
continuity equation for the entropy density in graphene in the absence
of temperature gradients (i.e., for ${\mu_\pm=\mu}$)
\begin{equation}
\label{ceenres}
\frac{\partial s}{\partial t} + \bs{\nabla}_{\bs{r}}\!\cdot\!(s\bs{u})
=
\frac{\bs{u}\!\cdot\!\bs{n}_{\bs{k}}}{T\tau_{\rm dis}}.
\end{equation}
The ideal hydrodynamic flow in the electronic system differs from its
usual counterpart by the presence of weak disorder scattering that is
all but unavoidable in any real solid.

\subsection{Relativistic hydrodynamics}

The linear spectrum of the elementary excitations in graphene,
Eq.~(\ref{eg}), suggests the possibility to observe relativistic
hydrodynamics in a solid-state laboratory experiments. In this
section, I compare the above ideal hydrodynamics in graphene to the
relativistic hydrodynamics in 2D \cite{har}.

\subsubsection{Hydrodynamic quantities in the relativistic theory}

The above ideal (Euler) hydrodynamics can be compared to the
standard relativistic hydrodynamics \cite{dau6,har} in $2+1$
dimensions, with $v_g$ playing the role of the speed of light. The
central quantity in this theory is the relativistic stress-energy
tensor
\begin{subequations}
\label{tik}
\begin{equation}
T^{ik}=wu^iu^k-pg^{ik},
\end{equation}
where $w$ and $p$ are the enthalpy and pressure, respectively, in the
local rest frame. $T^{ik}$ comprises the energy and momentum densities
as well as the $2\times2$ momentum flux tensor. Explicitly, the
individual components of $T^{ik}$ are as follows: the energy density
is given by $T^{00}$,
\begin{equation} 
\label{t00}
T^{00} = \frac{w}{1\!-\!u^2/v_g^2} - p;
\end{equation}
the momentum density is given by $T^{0\alpha}/v_g$ (in this section,
Greek indices refer to space and Roman -- to space-time),
\begin{equation} 
\label{t0a}
T^{0\alpha} = \frac{w u_\alpha}{v_g(1\!-\!u^2/v_g^2)};
\end{equation}
and the momentum flux density is given by $T^{\alpha\beta}$,
\begin{equation} 
\label{tba}
T^{\alpha\beta} = \frac{w u_\alpha u_\beta}{v_g^2(1\!-\!u^2/v_g^2)}+p\delta_{\alpha\beta}.
\end{equation}
\end{subequations}
The energy flux density is proportional to the momentum density and is
given by $v_gT^{0\alpha}$.

Resolving Eq.~(\ref{enth}) for the energy density $n_E$, I find
\begin{subequations}
\label{hqs}
\begin{equation}
\label{nerg}
n_{E} = \frac{3P}{1\!-\!u^2/v_g^2} - P.
\end{equation}
This expression is similar to Eq.~(\ref{t00}) since in the local rest
frame ${w=3p}$. However, unlike the pressure $p$ (defined in the local
rest frame), $P$ is the thermodynamic pressure calculated with the
local equilibrium distribution (\ref{le}) in the ``laboratory frame''
with $\bs{u}\ne0$.

The momentum density, $\bs{n}_{\bs{k}}$, can be expressed in the form
similar to Eq.~(\ref{t0a}) by combining Eqs.~(\ref{hjepiw}),
(\ref{nk}), and (\ref{enth}):
\begin{equation}
\label{nkrg}
\bs{n}^{(0)}_{\bs{k}} = \frac{W\bs{u}}{v_g^2} = \frac{3P\bs{u}}{v_g^2(1\!-\!u^2/v_g^2)}.
\end{equation}
Again, the difference between the two expressions is that between $p$
and $P$.

Finally, the momentum flux (\ref{hpi}) can be re-written similarly to
the relativistic expression (\ref{tba}) as well
\begin{equation}
\label{pirg}  
\Pi_{E,0}^{\alpha\beta}
= 
\frac{W}{3}\delta_{\alpha\beta} \!\left(\!1\!-\!\frac{u^2}{v_g^2}\right)
\!+\! 
\frac{W u_\alpha u_\beta}{v_g^2} 
=
P \delta_{\alpha\beta}
\!+\!
\frac{3Pu_\alpha u_\beta}{v_g^2(1\!-\!u^2/v_g^2)}.
\end{equation}
\end{subequations}
As a result, all three expressions (\ref{hqs}) are similar to their
relativistic counterparts (\ref{tik}), but are determined by the
pressure $P$ defined in the ``laboratory frame'', see Eq.~(\ref{p}),
instead of the ``relativistic'' pressure $p$ defined in the local rest
frame. On one hand, the reason $p$ is typically defined in the local
rest frame is simply the lack of a better definition: the standard
argument \cite{dau6} relies on the Pascal law that is valid only in
the local rest frame. On the other hand, the difference between the
two theories is significant: expressions (\ref{hqs}) cannot be
obtained from their rest-frame counterparts by the Lorentz
transformation. The reason for this is that the local equilibrium
distribution function (\ref{le}) is not relativistic. This can be
traced to the classical (i.e. non-relativistic) nature of the Coulomb
interaction in graphene that is ultimately responsible for
equilibration.

\subsubsection{Relativistic Euler equation}
\label{eulerd1}

Let me now compare the generalized Euler equation (\ref{eq0g}) to the
standard equations of relativistic hydrodynamics. These are encoded in
the relation \cite{dau6}
\begin{equation}
\label{rh1}
{\cal A}_i\equiv\frac{\partial T^k_i}{\partial x^k}=0.
\end{equation}

Substituting the stress-energy tensor of the ideal fluid (\ref{tik}),
one arrives at the equation \cite{dau6}
\begin{equation}
\label{rh2}
{\cal A}_i^{(0)}=u_i\frac{\partial (w u^k)}{\partial x^k} 
+ w u^k \frac{\partial u_i}{\partial x^k}
- \frac{\partial p}{\partial x^i}
=0.
\end{equation}
The relativistic generalization of the Euler equation is typically
obtained \cite{dau6} by projecting Eq.~(\ref{rh2}) on to the
direction orthogonal to $u^i$. This is achieved by considering the
combination
\begin{equation}
\label{proj0}
{\cal P}_i^{(0)}\equiv\left(\delta_i^k\!-\!u_i u^k\right) {\cal A}_k^{(0)} =
\frac{\partial T^k_i}{\partial x^k} - u_i u^k \frac{\partial T^l_k}{\partial x^l} = 0,
\end{equation}
which vanishes upon multiplication by $u^i$. Using Eq.~(\ref{rh2})
and the standard properties of the relativistic $3$-velocity (in 2D)
\begin{equation}
\label{3vp}
u^k u_k = 1, 
\qquad
u_k \frac{\partial u^k}{\partial x^l} = 0.
\end{equation}
one finds \cite{dau6}
\begin{equation}
\label{rh3}
{\cal P}_i^{(0)} = wu^k \frac{\partial u_i}{\partial x^k} - \frac{\partial p}{\partial x^i}
+ u_i u^k \frac{\partial p}{\partial x^k}=0.
\end{equation}
The Euler equation is contained in the spatial components of
Eq.~(\ref{rh3})
\begin{subequations}
\label{po}
\begin{equation}
\label{posp}
\bs{\cal P}^{(0)} \!\!=\! -\! \frac{wu_0}{v_g^2}
\!\left[\!\frac{\partial}{\partial t} \!+\! \bs{u}\!\cdot\!\bs{\nabla}\!\right]\! u_0 \bs{u}
\!-\!
\bs{\nabla} p
\!-\! \frac{u_0^2\bs{u}}{v_g^2} 
\!\left[\!\frac{\partial}{\partial t} \!+\! \bs{u}\!\cdot\!\bs{\nabla}\!\right]\! p
\!=\!0.
\end{equation}
This equation can be simplified with the help of the time component of
Eq.~(\ref{rh3}). Indeed, the time component of the left-hand side of
Eq.~(\ref{rh3}) reads
\begin{equation}
\label{pot}
{\cal P}^{(0)}_0=\frac{wu_0}{v_g}
\left[\frac{\partial}{\partial t}\!+\!\bs{u}\!\cdot\!\bs{\nabla}\right] u_0
-\frac{1}{v_g}\frac{\partial p}{\partial t}
+
\frac{u_0^2}{v_g} 
\left[\frac{\partial}{\partial t} \!+\! \bs{u}\!\cdot\!\bs{\nabla}\right]p=0.
\end{equation}
\end{subequations}
Combining Eqs.~(\ref{po}) as
\[
\bs{\cal P}^{(0)} + \frac{\bs{u}}{v_g}{\cal P}^{(0)}_0 
=
- \bs{\nabla} p 
-\frac{\bs{u}}{v_g^2}
\frac{\partial p}{\partial t}
-\frac{wu_0^2}{v_g^2}
\left[\frac{\partial}{\partial t} \!+\! \bs{u}\!\cdot\!\bs{\nabla}\right]\bs{u}
= 0,
\]
one finds the relativistic version of the Euler equation:
\begin{equation}
\label{reul}
\frac{w}{1\!-\!u^2/v_g^2}
\left[\frac{\partial}{\partial t} \!+\! \bs{u}\!\cdot\!\bs{\nabla}\right] \bs{u}
+
v_g^2\bs{\nabla} p +
\bs{u}\frac{\partial p}{\partial t} = 0.
\end{equation}
Similarly to the hydrodynamic quantities in Eqs.~(\ref{tik}), the
relativistic equation (\ref{reul}) differs from the generalized Euler
equation in graphene (\ref{eq0g}) in the absence of the
electromagnetic fields and disorder scattering insofar it contains the
rest frame pressure $p$ instead of the hydrodynamic pressure $P$.

Taking into account the electromagnetic fields, one replaces 
Eq.~(\ref{rh1}) by
\begin{subequations}
\label{fjt}
\begin{equation}
\label{fj}
\frac{\partial T^k_i}{\partial x^k} = \frac{e}{c}F_{ik}j^k.
\end{equation}
However, this would be wrong since the left-hand side transforms with
the Lorentz transformation with the velocity $v_g$, while the
right-hand side with $c\gg v_g$. The authors of Ref.~\cite{har}
suggested to rectify this issue by modifying the electromagnetic field
tensor, $F_{ik}$, and the $3$-current, $j^k$, such that the above
equation made sense. Indeed, replacing the textbook expressions by
\begin{equation}
\label{fikv}
F_{ik}=
\begin{pmatrix}
0 & (c/v_g) E_x & (c/v_g) E_y \cr
-(c/v_g) E_x & 0 & -B \cr
-(c/v_g) E_y & B & 0
\end{pmatrix},
\end{equation}
\begin{equation}
\label{jiv}
j^k = 
\begin{pmatrix}
v_g n, & \bs{j}
\end{pmatrix}
\end{equation}
\end{subequations}
one can recover Eq.~(\ref{cene1}) and the field dependent terms in
Eqs.~(\ref{cek1}) and (\ref{eq0g}).

Consider, however, Maxwell's equations
\[
\frac{\partial F^{ik}}{\partial x^k} = - \frac{4\pi}{c} j^i,
\quad
\epsilon^{iklm} \frac{\partial F_{lm}}{\partial x^k} = 0.
\]
Using the above expressions, one can directly verify, that the two
Maxwell's equations containing only electric and only magnetic fields
are preserved:
\[
\bs{\nabla}\!\times\!\bs{B}=0,
\qquad
\bs{\nabla}\!\cdot\!\bs{E}=4\pi\rho.
\]
However, the two remaining equations coupling the electric and
magnetic fields are clearly violated. Therefore, one should use the
relativistic approach with care.

\subsubsection{Thermodynamic quantities and entropy}

Combining the equation of motion (\ref{rh1}) with the
relativistic continuity equation
\begin{equation}
\label{rce1}
\frac{\partial (nu^i)}{\partial x^i}=0,
\end{equation}
one can derive the relativistic analog of the continuity equation for
the entropy density (\ref{ceenres}). Indeed, projecting
Eq.~(\ref{rh1}) onto the direction of the $3$-velocity,
\[
u^i\frac{\partial T^k_i}{\partial x^k} = 0
\quad\Rightarrow\quad
\frac{\partial (wu^k)}{\partial x^k} = u^k \frac{\partial p}{\partial x^k},
\]  
and using the thermodynamic (Gibbs-Duhem) relation
\[
w=\mu n + T {s},
\qquad
dp = nd\mu + sdT,
\]
where $s$ is the entropy density (in the rest frame), one finds
\[
\mu \frac{\partial (nu^k)}{\partial x^k} + T \frac{\partial (su^k)}{\partial x^k} = 0.
\]
The first term vanishes due to Eq.~(\ref{rce1}) and thus
\begin{equation}
\label{ceenr}
\frac{\partial (su^k)}{\partial x^k} = 0.
\end{equation}
Here  
\[
s u^k = (s, s\bs{u}).
\]
Thus the relativistic continuity relation has a similar form to the
continuity equation for the entropy in graphene (with the entropy
density $s$ and current $\bs{j}_S$ combined into a $3$-current $su^k$)
in the absence of disorder scattering. 

Similarly to all above cases of such correspondence, the entropy
density in the relativistic theory is defined in the local rest frame,
unlike the entropy density in graphene (\ref{enth}) which is defined
with the local equilibrium distribution of a moving electronic fluid
in the laboratory frame.

\section{Dissipative corrections to electronic hydrodynamics}

The above derivation of the generalized Euler equation (\ref{eq0g})
relies on the assumption of local equilibrium that is supposed to be
established by electron-electron collisions. The same scattering
processes are responsible for dissipation, i.e. irreversible charge
and momentum transfer from faster elements of the electronic fluid to
the slower ones. The general form of the dissipative corrections
follows from general arguments \cite{dau6}. Hydrodynamic quantities
are supposed to vary slowly over long distances, such that their
gradients should be small. Consequently, the dissipative correction to
the momentum flux tensor $\Pi_E^{\alpha\beta}$ should be linear in the
gradients of velocity. The specific form of the correction is governed
by rotational invariance and in 2D in the absence of external magnetic
field is given by
\begin{equation}
\label{dpi0}
\delta\Pi_E^{\alpha\beta} = - \eta (\nabla^\alpha u^\beta + \nabla^\beta u^\alpha
- \delta^{\alpha\beta}\bs{\nabla}\!\cdot\!\bs{u}) 
- \zeta \delta^{\alpha\beta}\bs{\nabla}\!\cdot\!\bs{u}.
\end{equation}
The shear and bulk viscosity coefficients ($\eta$ and $\zeta$,
respectively) can be found by a solution of the kinetic equation
\cite{dau10}. In the usual case of, e.g., a dilute gas, one can solve
the Boltzmann equation by means of the perturbative Chapman-Enskog
method \cite{chap,chap2,ensk1,ensk,bru}. Same results, although with a less
clear justification, may be obtained using the Grad method \cite{gra}.

Following the standard derivation of hydrodynamic equations from the
kinetic theory \cite{dau10,hydro1,har,luc}, I will look for the
dissipative corrections to the ideal hydrodynamics (\ref{eq0g}) within
linear response. The ``linear response'' solution should be obtained
by linearizing the collision integral in Eq.~(\ref{ke}) in the small
deviations from local equilibrium, ${\delta{f}=f-f^{(0)}}$, while
leaving only $f^{(0)}$ on the left-hand side:
\begin{equation}
\label{ke1}
{\cal L} f^{(0)} = {\rm St} [f].
\end{equation}
Note that by definition, ${\rm St}_{ee}[f^{(0)}]=0$.

Within the classic approach \cite{dau10}, one evaluates the Liouville's
operator on the left-hand side of the above expression explicitly
(with ${\bs{u}=0}$), then uses the ideal Euler equation and
thermodynamic relations to express the result in the form explicitly
containing the dissipative terms (again, with ${\bs{u}=0}$) as
``external forces''. The goal of such calculation is to find the
coefficients describing the dissipative corrections, i.e. viscosity
and thermal conductivity. This approach hinges on the fact that the
general form of the dissipative terms is known from symmetry arguments
(up to the coefficients).

In pure (disorder-free) graphene, the conserved current is the energy
current (since it is proportional to the momentum density), hence the
dissipative coefficients include viscosity and electrical
conductivity. My goal here is not only to determine these
coefficients, but also to establish the form of the dissipative
corrections. Therefore, instead of the direct evaluation of the
Liouville's operator, I will integrate the kinetic equation (\ref{ke1})
following \cite{hydro1}.

For Dirac fermions in graphene the solution of the kinetic equation is
simplified by the kinematic peculiarity of electron-electron
scattering known as the ``collinear scattering singularity''. For
Dirac quasiparticles moving along the same direction the energy and
momentum conservation laws coincide leading to a formal divergence of
the collision integral. Although the divergence is regularized by
dynamical screening, the resulting scale separation allows for a
nonperturbative solution.

\subsection{Collision integral due to electron-electron interaction}

The local equilibrium distribution function (\ref{le}) nullifies the
collision integral. Assuming that the external fields and other
perturbations lead to ``small'' deviations from local equilibrium, the
collision integral, ${\rm St}_{ee}[f]$, can be linearized in the
small, non-equilibrium correction to $f_{\lambda\bs{k}}^{(0)}$
\cite{dau10}
\begin{equation}
  \label{df}
  \delta f_{\lambda\bs{k}} = f_{\lambda\bs{k}} \!-\! f_{\lambda\bs{k}}^{(0)}
  = -T\frac{\partial f_{\lambda\bs{k}}^{(0)}}{\partial\epsilon_{\lambda\bs{k}}} h_{\lambda\bs{k}}
  = f_{\lambda\bs{k}}^{(0)}\left(1\!-\!f_{\lambda\bs{k}}^{(0)}\right) h_{\lambda\bs{k}}.
\end{equation}
The linearized collision integral can be written as \cite{dau10} (the
summation runs over all single-particle states up to the degeneracy
factor which is written down explicitly)
\begin{subequations} 
\label{collint}
\begin{equation}
{\rm St}_{ee}[f]\approx N\sum_{1,1',2'} W_{12,1'2'} f_{1}^{(0)}f_{2}^{(0)}
\left[1\!-\!f_{1'}^{(0)}\right]\left[1\!-\!f_{2'}^{(0)}\right]
\Big[h_{1'}\!+\!h_{2'}\!-\!h_1\!-\!h_2\Big],
\qquad
\sum_1 \equiv \sum_{\lambda_1} \int\frac{d^2k_1}{(2\pi)^2}.
\end{equation}
The transition probability $W_{12,1'2'}$ can be written using the
Fermi Golden Rule (e.g., neglecting interference effects \cite{zna})
\begin{equation}
\label{w0}
W_{12,34} \!=\! (2\pi)^3 |U|^2 \delta(\epsilon_1+\epsilon_2-\epsilon_{3}-\epsilon_{4})
\delta(\bs{k}_1+\bs{k}_2-\bs{k}_{3}-\bs{k}_{4}),
\end{equation}
\end{subequations}
where $U$ stands for the dynamically screened Coulomb interaction.

\subsection{Nonequilibrium correction to the distribution function}

The two $\delta$-functions in Eq.~(\ref{w0}) represent energy and
momentum conservation in an electron-electron ``collision''. For Dirac
fermions moving in the same direction they are identical and ${\rm
  St}_{ee}[f]$ diverges for a generic $h_{\lambda\bs{k}}$. There are
however three exceptions,
\[
h \propto \bs{k}, \bs{v}, \lambda\bs{v}.
\]
In the first case, the collision integral vanishes due to momentum
conservation, while in the other two the collision integral vanishes
for collinear particles. As a result, one can limit the mode expansion
of the nonequilibrium correction to the distribution function, $h$, to
the above three modes.

Adopting the ``three-mode approximation'', I can write the correction
$h$ in the form \cite{hydro1}
\begin{subequations}
\label{hs}
\begin{equation}
\label{hs0}
h_{\lambda\bs{k}} = \frac{\bs{v}_{\lambda\bs{k}}}{v_g}\sum_1^3 \phi_i \bs{h}^{(i)}
+ \frac{v^\alpha_{\lambda\bs{k}} v^\beta_{\lambda\bs{k}}}{v_g^2}  \sum_1^3 \phi_i h_{\alpha\beta}^{(i)} + \dots,
\end{equation}
where $\dots$ stands for higher-order tensors and the ``three modes''
are expressed by means of
\begin{equation}
\label{phis}
\phi_1 = 1, \quad \phi_2 = \lambda, \quad \phi_3 = \epsilon_{\lambda\bs{k}}/T.
\end{equation}
In accordance with the general strategy of evaluating the dissipative
corrections in the co-moving frame \cite{dau10}, all the coefficients
have to be considered in the limit $\bs{u}\rightarrow0$ (assuming they
are independent of velocity, at least for small enough $\bs{u}$). This
allows for a separate calculation of the vector and tensor quantities.

The coefficients $\bs{h}^{(i)}$ and $h_{\alpha\beta}^{(i)}$ in
Eq.~(\ref{hs0}) satisfy general constraints \cite{dau10} based on the
fact that electron-electron collisions do not alter conserved
thermodynamic quantities. To maintain momentum conservation, I should
set
\begin{equation}
\label{h30}
\bs{h}^{(3)}(\tau_{\rm dis}\rightarrow\infty)=0.
\end{equation} 
In the presence of weak disorder momentum is no longer conserved and
the energy current also acquires a dissipative correction. In this
case, I have to keep a nonzero $\bs{h}^{(3)}$ and then study the
(nontrivial) limit ${\tau_{\rm dis}\rightarrow\infty}$.

Now, to maintain conservation of the number of particles and energy I
set
\begin{equation}
\label{trh0}
{\rm Tr}\, h_{\alpha\beta}^{(i)}=0.
\end{equation}
\end{subequations}
The remaining coefficients can be determined by an explicit evaluation
of the corresponding macroscopic quantities \cite{hydro1}.

The macroscopic currents associated with the three modes $\phi_i$ are
the electric, imbalance, and energy currents. Using the nonequilibrium
distribution function in the definitions, Eqs.~(\ref{j}), (\ref{ji}),
and (\ref{je1}), I define the dissipative corrections (see also
\ref{ndflr})
\begin{equation}
\label{disjs}
\delta\bs{j} = N\sum_\lambda\!\int\!\frac{d^2k}{(2\pi)^2} \bs{v}_{\lambda\bs{k}}
\delta f_{\lambda\bs{k}},
\quad
\delta\bs{j}_I = N\sum_\lambda\lambda\!\int\!\frac{d^2k}{(2\pi)^2} \bs{v}_{\lambda\bs{k}}
\delta f_{\lambda\bs{k}},
\quad
\bs\delta{j}_E = N\sum_\lambda\lambda\!\int\!\frac{d^2k}{(2\pi)^2}
      \bs{v}_{\lambda\bs{k}}\epsilon_{\lambda\bs{k}}
\delta f_{\lambda\bs{k}}.
\end{equation}
Substituting Eq.~(\ref{hs0}) and evaluating the integrals (the tensor
terms do not contribute for $\bs{u}\rightarrow0$), I find the
following relation between the three corrections (\ref{disjs}) and the
vector coefficients $\bs{h}^{(i)}$
\begin{equation}
\label{djs}
\begin{pmatrix}
\delta\bs{j} \cr
\delta\bs{j}_I \cr
\delta\bs{j}_E/T
\end{pmatrix}
=
\frac{v_gT}{2} 
\widehat{M}_h\!
\begin{pmatrix}
\bs{h}^{(1)} \cr
\bs{h}^{(2)} \cr
\bs{h}^{(3)}
\end{pmatrix},
\qquad
\widehat{M}_h
=
\begin{pmatrix}
 \frac{\partial n}{\partial\mu} & \frac{\partial n_{I}}{\partial\mu} & \frac{2n}{T} \cr
 \frac{\partial n_{I}}{\partial\mu} & \frac{\partial n}{\partial\mu} & \frac{2n_{I}}{T} \cr
 \frac{2n}{T} & \frac{2n_{I}}{T} & \frac{3n_{E}}{T^2} \cr
\end{pmatrix},
\end{equation}
where the matrix elements of $\widehat{M}_h$ are expressed in terms of
the equilibrium densities (\ref{n0s}), (\ref{ne0}) and
compressibilities (\ref{dndmu}).

The tensor coefficients, $h_{\alpha\beta}^{(i)}$, in the second term
of the nonequilibrium correction (\ref{hs0}) are similarly related to
the dissipative corrections to the three macroscopic tensor quantities
(\ref{pab}), (\ref{piab}), and (\ref{piIab}) 
\begin{subequations}
\label{dpis}
\begin{equation}
\delta \Pi^{\alpha\beta} = N\sum_\lambda\!\int\!\frac{d^2k}{(2\pi)^2} 
v^\alpha_{\lambda\bs{k}}v^\beta_{\lambda\bs{k}} \delta f_{\lambda\bs{k}},
\quad
\delta \Pi_I^{\alpha\beta} = N\sum_\lambda\int\!\frac{d^2k}{(2\pi)^2}
\lambda v^\alpha_{\lambda\bs{k}}v^\beta_{\lambda\bs{k}} \delta f_{\lambda\bs{k}},
\end{equation}
\begin{equation}
\delta \Pi^{\alpha\beta}_E = \frac{1}{v_g^2} N\sum_\lambda\int\!\frac{d^2k}{(2\pi)^2} 
\epsilon_{\lambda\bs{k}} v^\alpha_{\lambda\bs{k}}v^\beta_{\lambda\bs{k}} \delta f_{\lambda\bs{k}}.
\end{equation}
\end{subequations}
Substituting the distribution function (\ref{hs0}), one finds
essentially the same integrals as in the case of the currents (now
only the tensor part of the nonequilibrium correction yields a nonzero
contribution) such that
\begin{equation}
  \label{m2}
  \begin{pmatrix}
    \delta \Pi^{\alpha\beta} /v_g^2 \cr
    \delta \Pi_I^{\alpha\beta} /v_g^2 \cr
    \delta \Pi_E^{\alpha\beta}/T
  \end{pmatrix}
  =
  \frac{T}{4}\widehat{M}_h
  \begin{pmatrix}
    h^{(1)}_{\alpha\beta}\cr
    h^{(2)}_{\alpha\beta}\cr
    h^{(3)}_{\alpha\beta}
  \end{pmatrix}.
\end{equation}

\subsection{Electrical conductivity}

The relation between the coefficients in the nonequilibrium
distribution function (\ref{hs0}) and the macroscopic currents,
Eq.~(\ref{djs}), suggests the following method of solving the
linearized kinetic equation (\ref{ke1}). Integrating the kinetic
equation, one obtains equations for the currents. Then using the
relation (\ref{djs}) one finds dissipative corrections to the currents
as linear functions of external fields.

\subsubsection{Macroscopic equation for the electric current}

The equation for the electric current is obtained by multiplying the
kinetic equation (\ref{ke}) by the velocity and integrating over all
single-particle states [cf. Eq.~(\ref{j})]. The resulting equation
will have the form
\begin{equation}
\label{jeq0}
N\!\sum_\lambda\!\!\int\!\!\frac{d^2k}{(2\pi)^2} \bs{v}_{\lambda\bs{k}} 
{\cal L}\Big|_{\bs{B}=0} f^{(0)}_{\lambda\bs{k}}
+
\frac{e}{c} N\!\sum_\lambda\!\!\int\!\!\frac{d^2k}{(2\pi)^2} \bs{v}_{\lambda\bs{k}}
\left( \left[ \bs{v}_{\lambda\bs{k}}\!\times\!\bs{B} \right]\!\cdot\!\bs{\nabla}_{\bs{k}} f \right)
= \bs{\cal I}_1\left[f\right]
\equiv N\!\sum_\lambda\!\!\int\!\!\frac{d^2k}{(2\pi)^2} \bs{v}_{\lambda\bs{k}}{\rm St}\left[f\right],
\end{equation}
where $\bs{\cal I}_1\left[f\right]$ is the integrated collision
integral and the Lorentz term is not treated within linear response.
Evaluating the integrals on the left-hand side of Eq.~(\ref{jeq0}),
one finds
\begin{equation}
\label{lj0l}
\frac{\partial \left(n u^\alpha\right)}{\partial t}
+
\nabla^\beta \Pi^{\alpha\beta}
-
eE^\beta \sum_\lambda \frac{\partial}{\partial \mu_\lambda}
\left[\Pi^{\alpha\beta}_{\lambda}\!-\!j^\alpha_{\lambda}u^\beta\right]
+
\omega_B \epsilon^{\alpha\beta\gamma} e_B^\beta {\cal K}^\gamma
= {\cal I}^\alpha_1\left[f\right],
\qquad
\omega_B = \frac{eBv_g^2}{c{\cal T}},
\end{equation}
where $\bs{e}_B$ is the unit vector in the direction of $\bs{B}$, the
tensors $\Pi^{\alpha\beta}$ and $\Pi_\lambda^{\alpha\beta}$ are
defined in Eq.~(\ref{piab}), the ``band currents'', $\bs{j}_\lambda$
are defined in Eqs.~(\ref{js}), $\omega_B$ is the generalized
cyclotron frequency, ${\cal T}$ is defined in Eq.~(\ref{t}), and the
vector quantity $\bs{\cal K}$ defined as
\begin{equation}
\label{lj04}
\bs{\cal K}={\cal T}N\sum_\lambda \int\!\frac{d^2k}{(2\pi)^2}
\frac{\bs{k}}{k^2} f_{\lambda\bs{k}}.
\end{equation}
has dimensions of the current.

Evaluating the integrated collision integral and the vector ${\cal K}$
using the nonequilibrium distribution function (\ref{hs0}) yields a
linear function of the coefficients $\bs{h}^{(i)}$. Details of the
calculation are relegated to \ref{ciceqs} and \ref{kvecs}. The result
is summarized below together with the equations for the two other
macroscopic currents.

\subsubsection{Macroscopic equation for the imbalance current}

The equation for the imbalance current is obtained similarly to
Eq.~(\ref{lj0l4B}): one multiplies the kinetic equation by
$\lambda\bs{v}_{\lambda\bs{k}}$ and integrates over all
single-particle states [see the definition (\ref{ji})]
\begin{equation}
\label{jIeq0}  N\!\sum_\lambda\!\!\int\!\!\!\frac{d^2k}{(2\pi)^2} 
\lambda\bs{v}_{\lambda\bs{k}} {\cal L}\Big|_{\bs{B}=0} f^{(0)}_{\lambda\bs{k}}
+
\frac{e}{c} N\!\sum_\lambda\!\lambda\!\int\!\!\frac{d^2k}{(2\pi)^2} \bs{v}_{\lambda\bs{k}}
\left( \left[ \bs{v}_{\lambda\bs{k}}\!\times\!\bs{B} \right]\!\cdot\!\bs{\nabla}_{\bs{k}} f \right)
\!=\! 
\bs{\cal I}_2\!\left[f\right]  
\!\equiv\! N\!\sum_\lambda\!\!\int\!\!\!\frac{d^2k}{(2\pi)^2} \lambda\bs{v}_{\lambda\bs{k}}{\rm St}[f],
\end{equation}
where $\bs{\cal I}_2\left[f\right]$ is the integrated collision
integral. Evaluating the integrals, I find an equation similar to
Eq.~(\ref{lj0l})
\begin{equation}
\label{ljI0l}
\frac{\partial \left(n_I u^\alpha\right)}{\partial t}
+
\nabla^\beta \Pi_I^{\alpha\beta}
-
eE^\beta \sum_\lambda\lambda \frac{\partial}{\partial \mu_\lambda}
\left[\Pi^{\alpha\beta}_{\lambda}\!-\!j^\alpha_{\lambda}u^\beta\right]
+
\omega_B \epsilon^{\alpha\beta\gamma} e_B^\beta {\cal K}_I^\gamma
= {\cal I}^\alpha_2\left[f\right],
\end{equation}
where
\begin{equation}
\label{ljI04}
\bs{\cal K}_I={\cal T}N\sum_\lambda \lambda\!\int\!\frac{d^2k}{(2\pi)^2}
\frac{\bs{k}}{k^2} f_{\lambda\bs{k}}.
\end{equation}
The integrated collision integral $\bs{\cal I}_2$ and the vector
$\bs{\cal K}_I$ are calculated in \ref{ciceqs} and \ref{kvecs},
respectively.

\subsubsection{Macroscopic equation for the energy current}

The equation for the energy current is given by Eq.~(\ref{cek1}),
multiplied by $v_g^2$. Substituting the ideal quantities into the
left-hand side (with the exception of the Lorentz term), one
finds 
\begin{equation}
\label{cek2}
\partial_t \bs{j}_E + \frac{1}{2}v_g^2\bs{\nabla}_{\bs{r}} n_E
- e n v_g^2 \bs{E} - v_g^2 \frac{e}{c} \left[\bs{j}\!\times\!\bs{B}\right] =
- \frac{\bs{j}_E}{\tau_{\rm dis}}.
\end{equation}
This equation differs from Eq.~(\ref{cek1}) and hence from the Euler
equation (\ref{eq0g}) by the fact that the electric current in the
Lorentz term and the energy current on the right-hand side are total
currents including the dissipative corrections. Adopting the standard
iterative method \cite{dau10} of the derivation of the dissipative
corrections, one has to separate the terms in Eq.~(\ref{cek2}) forming
the Euler equation and the terms containing $\delta\bs{j}$ and
$\delta\bs{j}_E$. Assuming the validity of the Euler equation as the
zeroth iteration, this leaves one with the following relation between
the dissipative corrections
\begin{equation}
\label{ljE}
v_g^2\frac{e}{c} \delta\bs{j}\!\times\!\bs{B} = \frac{\delta\bs{j}_E}{\tau_{\rm dis}}
\qquad\Rightarrow\qquad
\omega_B \frac{{\cal T}}{T}\delta\bs{j}\!\times\!\bs{e}_B = 
\frac{1}{\tau_{\rm dis}}\frac{\delta\bs{j}_E}{T}.
\end{equation}
The right-hand side yields the explicit form of the integrated
collision integral (due to disorder only since the electron-electron
interaction conserves momentum).

The simple form of Eq.~(\ref{ljE}) has a simple physical
meaning. Since the energy current is proportional to the momentum
density, it cannot be relaxed by electron-electron interaction (which
conserves momentum). Consequently, a steady state cannot be achieved
without disorder scattering contradicting the use of time-independent
corrections to macroscopic currents (\ref{disjs}).

\subsubsection{Dissipative corrections to quasiparticle currents}

The three equations (\ref{lj0l}), (\ref{ljI0l}), and (\ref{cek2})
coincide with the macroscopic linear response equations derived in
Ref.~\cite{hydro0}. These equations are (at least, in principle)
sufficient for describing traditional linear response transport in
graphene and are valid even for relatively strong disorder, where the
hydrodynamic approach is invalid. At the same time, for weak disorder,
i.e., within the applicability region of the hydrodynamic theory, the
latter provides a significant generalization of the linear response
theory allowing for a description of the collective motion of a
strongly interacting fluid.

Applying the above iterative approach of the derivation of the
dissipative corrections to the ideal Euler hydrodynamics, I now
simplify Eqs.~(\ref{lj0l}) and (\ref{ljI0l}) assuming the validity of
the Euler equation for the ideal quantities in their respective
left-hand sides. In particular, the Euler equation (\ref{eq0g})
can be used to express the time derivative of the velocity in terms of
the pressure gradient and electromagnetic fields. The pressure
gradient can be expressed in terms of the gradient of $n_E$ using the
equation of state (\ref{pg}). Finally, density gradients can be
expressed in terms of gradients of temperature and chemical
potential. As a result, I arrive at the equations
\begin{equation}
\label{lj0l4B}
\left[
\frac{2n^2}{3n_{E}}
-
\frac{1}{2}  \frac{\partial n}{\partial \mu}
\right]\!
\left[ e\bs{E}-T\,\bs{\nabla}\frac{\mu}{T}+\frac{e}{c}\bs{u}\!\times\!\bs{B}\right]
-
\left[\frac{2n n_{I}}{3n_{E}}-\frac{1}{2}\frac{\partial n_{I}}{\partial \mu}
\right]T\,\bs{\nabla}\frac{\mu_I}{T}
= \frac{1}{v_g^2}
\left(\bs{\cal I}_1^{ee} - \frac{\delta\bs{j}}{\tau_{\rm dis}}
\!-\!\omega_B \bs{e}_B\!\times\!\delta\bs{\cal K}
\right)\!,
\end{equation}
\begin{equation}
\label{ljI0l4B}
\left[
\frac{2n n_{I}}{3n_{E}}
-
\frac{1}{2}  \frac{\partial n_{I}}{\partial \mu}
\right]\!
\left[ e\bs{E}-T\,\bs{\nabla}\frac{\mu}{T}+\frac{e}{c}\bs{u}\!\times\!\bs{B}\right]
-
\left[\frac{2n^2_{I}}{3n_{E}}-\frac{1}{2}\frac{\partial n}{\partial \mu}
\right]T\,\bs{\nabla}\frac{\mu_I}{T}
= 
\frac{1}{v_g^2}
\left(\bs{\cal I}_2^{ee} - \frac{\delta\bs{j}_I}{\tau_{\rm dis}} 
- \omega_B \bs{e}_B\!\times\!\delta\bs{\cal K}_I
\right)\!.
\end{equation}

Combining the integrated equations for macroscopic currents
(\ref{lj0l4B}), (\ref{ljI0l4B}), and (\ref{ljE}) with the integrated
collision integrals (see \ref{ciceqs}) and Lorentz terms (see
\ref{kvecs}), I obtain the final set of linear equations for the
dissipative corrections to macroscopic currents (\ref{disjs})
\begin{subequations}
\label{djeqsB}
\begin{equation}
\label{djeq1}
\widehat{M}_n \!
\begin{pmatrix}
e\bs{E}\!-\!T\,\bs{\nabla}\displaystyle\frac{\mu}{T}\!+\!\frac{e}{c}\bs{u}\!\times\!\bs{B} \cr
T\,\bs{\nabla}\displaystyle\frac{\mu_I}{T} \cr
0
\end{pmatrix}\!
=
-\frac{1}{v_g^2}\!
\left[
\frac{\partial n}{\partial \mu}
\widehat{T}_m
\widehat{M}_h^{-1}
\!+\!
\frac{1}{\tau_{\rm dis}}
\widehat{\mathbb{1}}
\right]\!\!
\begin{pmatrix}
  \delta\bs{j} \cr
  \delta\bs{j}_I \cr
  \delta\bs{j}_E/T
\end{pmatrix}\!
-
\frac{\omega_B}{v_g^2}\frac{\partial n}{\partial \mu}
\widehat{\textswab{M}}_K
\widehat{M}_h^{-1}
\bs{e}_B\!\times\!
\begin{pmatrix}
  \delta\bs{j} \cr
  \delta\bs{j}_I \cr
  \delta\bs{j}_E/T 
\end{pmatrix}\!,
\end{equation}
where I define the following matrices (and their dimensionless
counterparts)
\begin{equation}
\label{mdim}
\widehat{M}_n 
\!=\! 
\begin{pmatrix}
\frac{2n^2}{3n_{E}}\!-\!\frac{1}{2}\frac{\partial n}{\partial \mu} &
-\frac{2n n_{I}}{3n_{E}}\!+\!\frac{1}{2}\frac{\partial n_{I}}{\partial \mu} & 0\cr
\frac{2n n_{I}}{3n_{E}}\!-\!\frac{1}{2}\frac{\partial n_{I}}{\partial \mu} &
-\frac{2n_{I}^2}{3n_{E}}\!+\!\frac{1}{2}\frac{\partial n}{\partial \mu} & 0\cr
0 & 0 & 0
\end{pmatrix}\!
=
- \frac{1}{2}\frac{\partial n}{\partial \mu} \widehat{\textswab{M}}_n,
\quad
\widehat{T}_m\!=\!
\begin{pmatrix}
  \tau_{11}^{-1} & \tau_{12}^{-1} & 0 \cr
  \tau_{12}^{-1} & \tau_{22}^{-1} & 0 \cr
  0 & 0 & 0
\end{pmatrix}
=
\frac{\alpha_g^2}{8\pi}\frac{NT^2}{\cal T} \widehat{\textswab{T}}.
\end{equation}
The matrix $\widehat{M}_n$ describes the left-hand sides of
Eqs.~(\ref{lj0l4B}) and (\ref{ljI0l4B}), while the matrix
$\widehat{T}_m$ comprises the ``scattering rates'' appearing in the
integrated collision integrals (\ref{li1r}). These two terms determine
the dissipative corrections $\delta\bs{j}$ and $\delta\bs{j}_I$ in the
absence of disorder and magnetic fields \cite{hydro1}. The second term
on the right-hand side in Eq.~(\ref{djeq1}) describes the effect of
disorder scattering. In the absence of the magnetic field disorder
scattering yields only a small correction to the effect of
electron-electron interaction represented by $\widehat{T}_m$. In the
presence of the magnetic field the role of disorder is more
pronounced: it is necessary to establish the steady state in the
system as follows from Eq.~(\ref{ljE}). The effect of the magnetic
field is described by the vectors ${\cal K}$ and ${\cal K}_I$, which
are linear combinations of the dissipative corrections. The
coefficients in these combinations, as well as in Eq.~(\ref{ljE}),
form the matrix (hereafter I consider the standard case
${\mu_\pm=\mu}$ or ${\mu_I=0}$)
\begin{equation}
\label{mk}
\widehat{\textswab{M}}_K
=
\begin{pmatrix}
\tanh\frac{x}{2} & 1 & \frac{\cal T}{T} \cr
1 & \tanh\frac{x}{2} & x \cr
\frac{\cal T}{T} & x & 2\tilde{n}
\end{pmatrix}.
\end{equation}
Introducing the dimensionless counterpart of the matrix $\widehat{M}_h$
\begin{equation}
\label{mhdim}
\widehat{M}_h = \frac{\partial n_{0}}{\partial \mu} \widehat{\textswab{M}}_h,
\qquad
\widehat{\textswab{M}}_h=
\begin{pmatrix}
1 & \frac{xT}{\cal T} & 2\tilde{n} \frac{T}{\cal T} \cr
\frac{xT}{\cal T} & 1 & \left[x^2\!+\!\frac{\pi^2}{3}\right]\frac{T}{\cal T} \cr
2\tilde{n} \frac{T}{\cal T} & \left[x^2\!+\!\frac{\pi^2}{3}\right]\frac{T}{\cal T} &
6\tilde{n}_E \frac{T}{\cal T}
\end{pmatrix},
\end{equation}
I re-write Eq.~(\ref{djeq1}) in the form
\begin{eqnarray}
\label{djeq2}
\widehat{\textswab{M}}_n
\begin{pmatrix}
  e\bs{E}\!-\!T\,\bs{\nabla}\displaystyle\frac{\mu}{T}\!+\!\frac{e}{c}\bs{u}\!\times\!\bs{B} \cr
  0 \cr
  0
\end{pmatrix}
=
\frac{\alpha_g^2T^2}{2{\cal T}^2}
\left[
\widehat{\textswab{T}}
+
\frac{1}{t_{\rm d}}
\widehat{\textswab{M}}_h
\right]
\widehat{\textswab{M}}_h^{-1}
\!
\begin{pmatrix}
  \delta\bs{j} \cr
  \delta\bs{j}_I \cr
  \delta\bs{j}_E/T 
\end{pmatrix}
+
\pi
\frac{\omega_B}{\cal T}
\widehat{\textswab{M}}_K
\widehat{\textswab{M}}_h^{-1}
\bs{e}_B\!\times\!
\begin{pmatrix}
  \delta\bs{j} \cr
  \delta\bs{j}_I \cr
  \delta\bs{j}_E/T 
\end{pmatrix}\!,
\end{eqnarray}
\end{subequations}
where $t_{\rm d}$ is the dimensionless impurity scattering time
defined similarly to the way the matrix $\widehat{\textswab{T}}$ is defined.

In general,the ${6\!\times\!6}$ matrix on the right-hand side of
Eq.~(\ref{djeq2}) may be inverted as follows. Introducing the
short-hand notations,
\begin{equation}
\label{sxx}
\widehat{\textswab{S}}_{xx}
=
\frac{\alpha_g^2T^2}{2{\cal T}^2}
\left[
\widehat{\textswab{T}}
+
\frac{1}{t_{\rm d}}
\widehat{\textswab{M}}_h
\right],
\quad
\widehat{\textswab{S}}_{xy}
=
\pi\frac{\omega_B}{\cal T}
\widehat{\textswab{M}}_K,
\quad
\textswab{E}
=
\begin{pmatrix}
  e\bs{E}\!-\!T\,\bs{\nabla}\displaystyle\frac{\mu}{T}\!+\!\frac{e}{c}\bs{u}\!\times\!\bs{B} \cr
  0 \cr
  0
\end{pmatrix},
\quad
\textswab{h}
=
\widehat{\textswab{M}}_h^{-1}
\begin{pmatrix}
  \delta\bs{j} \cr
  \delta\bs{j}_I \cr
  \delta\bs{j}_E/T 
\end{pmatrix},
\end{equation}
I can re-write Eq.~(\ref{djeq2}) as follows
\[
\widehat{\textswab{M}}_n\textswab{E}
=
\widehat{\textswab{S}}_{xx}\textswab{h}+\widehat{\textswab{S}}_{xy}\bs{e}_B\!\times\!\textswab{h}.
\]
Multiplying this equation by $\bs{e}_B$ (using the fact that this
vector product acts in the position space and hence commutes with all
the matrices), I obtain
\[
\widehat{\textswab{M}}_n\bs{e}_B\!\times\!\textswab{E}
=
\widehat{\textswab{S}}_{xx}\bs{e}_B\!\times\!\textswab{h}-\widehat{\textswab{S}}_{xy}\textswab{h}.
\]
The two equations can now be solved as a usual system of two linear
equations with the only difference, that the coefficients are now
matrices that do not commute. Hence, one has to keep track of the
order in which they are multiplied. The resulting solution has the
form
\begin{eqnarray}
\label{sigmagen}
&&
\begin{pmatrix}
  \delta\bs{j} \cr
  \delta\bs{j}_I \cr
  \delta\bs{j}_E/T 
\end{pmatrix}
=
\widehat{\textswab{M}}_h
\left(
1\!+\!\textswab{S}_{xx}^{-1}\textswab{S}_{xy}\textswab{S}_{xx}^{-1}\textswab{S}_{xy}
\right)^{-1}
\textswab{S}_{xx}^{-1}\widehat{\textswab{M}}_n
\begin{pmatrix}
  e\bs{E}\!-\!T\,\bs{\nabla}\frac{\mu}{T}\!+\!\frac{e}{c}\bs{u}\!\times\!\bs{B} \cr
  0 \cr
  0
\end{pmatrix}
\\
&&
\nonumber\\
&&
\qquad\qquad\qquad\qquad\quad
-\,
\widehat{\textswab{M}}_h
\left(
1\!+\!\textswab{S}_{xx}^{-1}\textswab{S}_{xy}\textswab{S}_{xx}^{-1}\textswab{S}_{xy}
\right)^{-1}
\textswab{S}_{xx}^{-1}\textswab{S}_{xy}\textswab{S}_{xx}^{-1}
\widehat{\textswab{M}}_n
\bs{e}_B\!\times\!
\begin{pmatrix}
  e\bs{E}\!-\!T\,\bs{\nabla}\frac{\mu}{T}\!+\!\frac{e}{c}\bs{u}\!\times\!\bs{B} \cr
  0 \cr
  0
\end{pmatrix},
\nonumber
\end{eqnarray}
which reminds one of the standard form of magnetoconductivity in the
Drude theory
\[
\sigma_{xx}=\frac{\sigma_D}{1\!+\!\omega_c^2\tau^2},
\qquad
\sigma_{xy}=\frac{\omega_c\tau\sigma_D}{1\!+\!\omega_c^2\tau^2}.
\]

The result (\ref{sigmagen}) expresses the dissipative corrections to
the macroscopic currents in the system as a function of the electric
field (more precisely, of the gradient of the electrochemical
potential) defining the dissipative coefficients in analogy with the
thermal conductivity in the traditional hydrodynamics \cite{dau6}.

\subsection{Viscosity}

Within the usual hydrodynamics \cite{dau6}, shear and bulk viscosities
are defined as the coefficients in the leading term in the gradient
expansion of the dissipative correction to the momentum flux tensor,
see Eq.~(\ref{dpi0}). In this section, I establish the form of this
correction in graphene following the same steps leading to the
corrections to the quasiparticle currents, Eq.~(\ref{sigmagen}).
Because I am now looking for corrections to a tensor quantity, the
second term in the nonequilibrium distribution function (\ref{hs0}) is
going to contribute.

\subsubsection{Macroscopic equations for tensor quantities}

Although the viscosity is defined though the dissipative correction to
only one macroscopic tensor quantity, the momentum flux tensor
$\Pi_E^{\alpha\beta}$, the three-mode approximation adopted in this
paper requires one to consider equations determining the three
macroscopic tensors: $\Pi^{\alpha\beta}$, $\Pi_I^{\alpha\beta}$, and
$\Pi_E^{\alpha\beta}$. Similarly to the equations for the corresponding currents, these
equations can be obtained by multiplying the kinetic equation by
$v^\alpha v^\beta$, $\lambda v^\alpha v^\beta$, and $\epsilon v^\alpha
v^\beta/T$ (respectively) and integrating over all states. The direct
integration yields the three equations (where the external electric
field is set to zero since I am looking for viscosity as a function
of magnetic field only).
\begin{subequations}
\label{teneqs}
\begin{equation}
\label{schaeq1B}
\frac{\partial \Pi^{\alpha\beta}}{\partial t}
+
\nabla^\gamma \Upsilon^{\alpha\beta\gamma}
=
{\cal I}^{\alpha\beta}_1
-
\omega_B \left[\epsilon^{\alpha ji} e_B^j \Xi^{i\beta}\!+\!\epsilon^{\beta ji} e_B^j \Xi^{i\alpha}\right]\!.
\end{equation}
\begin{equation}
\label{schaIeq1B}
\frac{\partial \Pi^{\alpha\beta}_{I}}{\partial t}
+
\nabla^\gamma \Upsilon^{\alpha\beta\gamma}_{I}
=
{\cal I}_2^{\alpha\beta}
-
\omega_B \left[\epsilon^{\alpha ji} e_B^j \Xi_I^{i\beta}
\!+\!\epsilon^{\beta ji} e_B^j \Xi_I^{i\alpha}\right]\!,
\end{equation}
\begin{equation}
\label{schaEeq1B}
\frac{v_g^2}{T}\frac{\partial \Pi^{\alpha\beta}_{E}}{\partial t}
+
\nabla^\gamma \Upsilon^{\alpha\beta\gamma}_{E}
=
{\cal I}_3^{\alpha\beta}
-
\omega_B \left[\epsilon^{\alpha ji} e_B^j \Xi_E^{i\beta}
\!+\!\epsilon^{\beta ji} e_B^j \Xi_E^{i\alpha}\right]\!.
\end{equation}
\end{subequations}

The third-rank tensors $\Upsilon^{\alpha\beta\gamma}$,
$\Upsilon^{\alpha\beta\gamma}_I$, and $\Upsilon^{\alpha\beta\gamma}_E$
appear in the integrated equations (\ref{teneqs}) in the same way as
the second-rank tensors $\Pi^{\alpha\beta}$, $\Pi_I^{\alpha\beta}$,
and $\Pi_E^{\alpha\beta}$ appear in the integrated equations for
the macroscopic currents. Similarly to the evaluation of the left-hand
sides of the integrated equations for macroscopic currents, these
quantities have to be computed with the local equilibrium distribution
function (\ref{le}). In the limit ${\bs{u}\rightarrow0}$ these
quantities are linear in $\bs{u}$ and would be discarded if it were
not for the fact that Eqs.~(\ref{teneqs}) contain only gradients of
these quantities. The straightforward calculation yields
\begin{subequations}
\label{ys}
\begin{equation}
\label{y1}
\Upsilon^{\alpha\beta\gamma} =
N\sum_\lambda\int\!\frac{d^2k}{(2\pi)^2}
v^\alpha_{\lambda\bs{k}}v^\beta_{\lambda\bs{k}}v^\gamma_{\gamma\bs{k}} f^{(0)}_{\lambda\bs{k}}
\quad\underset{\bs{u}\rightarrow0}{\longrightarrow}\quad
\frac{1}{4} v_g^2 n \!\left(
u^\alpha\delta^{\beta\gamma}\!+\!u^\beta\delta^{\alpha\gamma}\!+\!u^\gamma\delta^{\alpha\beta}
\right),
\end{equation}
\begin{equation}
\label{y2}
\Upsilon^{\alpha\beta\gamma}_{I} =
N\sum_\lambda\int\!\frac{d^2k}{(2\pi)^2}\lambda 
v^\alpha_{\lambda\bs{k}}v^\beta_{\lambda\bs{k}}v^\beta_{\gamma\bs{k}} f^{(0)}_{\lambda\bs{k}}
\quad\underset{\bs{u}\rightarrow0}{\longrightarrow}\quad
\frac{v_g^2}{4}  n_{I} \!\left(
u^\alpha\delta^{\beta\gamma}\!+\!u^\beta\delta^{\alpha\gamma}\!+\!u^\gamma\delta^{\alpha\beta}
\right),
\end{equation}
\begin{equation}
\label{y3}
\Upsilon^{\alpha\beta\gamma}_{E} =
\frac{N}{T}
\sum_\lambda\int\!\frac{d^2k}{(2\pi)^2}\epsilon_{\lambda\bs{k}}
v^\alpha_{\lambda\bs{k}}v^\beta_{\lambda\bs{k}}v^\gamma_{\gamma\bs{k}} f^{(0)}_{\lambda\bs{k}}
\quad\underset{\bs{u}\rightarrow0}{\longrightarrow}\quad
\frac{3}{8} \frac{v_g^2}{T}n_{E}\left(
u^\alpha\delta^{\beta\gamma}\!+\!u^\beta\delta^{\alpha\gamma}\!+\!u^\gamma\delta^{\alpha\beta}
\right).
\end{equation}
\end{subequations}
Now, in the limit ${\bs{u}\rightarrow0}$ the second-rank tensors are
proportional to the corresponding densities, see
Eq.~(\ref{hpi}). Using the continuity equations to express the time
derivatives of densities in terms of gradients similarly to the
transformations used to derive Eqs.~(\ref{lj0l4B}) and
(\ref{ljI0l4B}), one can simplify the left-hand sides of
Eqs.~(\ref{teneqs}). The result can be expressed in vector form as
\begin{equation}
\label{schaeqB}
\left(
\nabla^\alpha u^\beta+\nabla^\beta u^\alpha-\delta^{\alpha\beta}\bs{\nabla}\!\cdot\!\bs{u}
\right)
\begin{pmatrix}
n \cr
n_{I} \cr
3n_{E}/(2T)
\end{pmatrix}
= 
\frac{4}{v_g^2}
\begin{pmatrix}
{\cal I}_1^{\alpha\beta}\!-\!\omega_B 
\left(\epsilon^{\alpha ji} e_B^j \Xi^{i\beta}\!+\!\epsilon^{\beta ji} e_B^j \Xi^{i\alpha}\right)  \cr
{\cal I}_2^{\alpha\beta}\!-\!\omega_B 
\left(\epsilon^{\alpha ji} e_B^j \Xi_I^{i\beta}\!+\!\epsilon^{\beta ji} e_B^j \Xi_I^{i\alpha}\right)  \cr
{\cal I}_3^{\alpha\beta}\!-\!\omega_B 
\left(\epsilon^{\alpha ji} e_B^j \Xi_E^{i\beta}\!+\!\epsilon^{\beta ji} e_B^j \Xi_E^{i\alpha}\right)  \cr
\end{pmatrix}\!.
\end{equation}

Comparing the left-hand side of Eq.~(\ref{schaeqB}) to the definition
(\ref{dpi0}), I can already conclude that the bulk viscosity in
graphene vanishes (at least within the approximations adopted in this
paper).

The integrated collision integrals ${\cal I}_j^{\alpha\beta}$ are
discussed in \ref{citeqs}. The integrated Lorentz terms contain the
tensors $\Xi^{i\alpha}$, $\Xi_I^{i\alpha}$, and $\Xi_E^{i\alpha}$ in
analogy with the vectors ${\cal K}$ and ${\cal K}_I$ in the above
conductivity calculation. The discussion of these tensors is relegated
to \ref{ksitens}. Combining them in a vector in the ``mode space'', I
find
\begin{equation}
\label{xivec}
\begin{pmatrix}
\Xi^{i\beta} \cr
\Xi^{i\beta}_I \cr
\Xi^{i\beta}_E
\end{pmatrix}
=
\frac{v_g^2T}{4} \frac{\partial n}{\partial\mu} \,
\widehat{\textswab{M}}_K
\begin{pmatrix}
h^{(1)}_{i\beta} \cr
h^{(2)}_{i\beta} \cr
h^{(3)}_{i\beta}
\end{pmatrix}
=
\widehat{\textswab{M}}_K
\widehat{\textswab{M}}_h^{-1}
\begin{pmatrix}
\delta\Pi^{i\beta} \cr
\delta\Pi^{i\beta}_I \cr
v_g^2\delta\Pi^{i\beta}_E /T
\end{pmatrix}\!,
\end{equation}
where the coefficients in Eq.~(\ref{ksi1}) are combined into the
matrix (\ref{mk}). Note, that according to the definitions
(\ref{ksidef}) the matrices $\Xi^{i\beta}_n$ are symmetric, but not
necessarily traceless. However, the matrices $h^{(n)}_{i\beta}$ are
traceless, see Eq.~(\ref{trh0}), hence the matrices $\Xi^{i\beta}_n$
are traceless as well.

Traceless, symmetric, $2\times2$ matrices contain only two independent
elements. Consequently, the last two terms in Eq.~(\ref{schaeqB}) must
be related to each other [in other words, Eq.~(\ref{schaeqB}) as a
system of two linear equations for the two matrix elements of
$\Xi^{i\beta}_n$]. Indeed, evaluating the spatial components
explicitly, I find
\[
\epsilon^{\alpha ji} e_B^j \Xi^{i\beta}
=
\begin{pmatrix}
-\Xi^{yx} & -\Xi^{yy} \cr
\Xi^{xx} & \Xi^{xy}
\end{pmatrix}\!,
\qquad
\epsilon^{\beta ji} e_B^j \Xi^{i\alpha}
=
\begin{pmatrix}
-\Xi^{yx} & \Xi^{xx} \cr
-\Xi^{yy} & \Xi^{xy}
\end{pmatrix}\!,
\]
which are identical, since 
\[
{\rm Tr}\;\Xi^{ij} = \Xi^{xx} + \Xi^{yy} = 0
\qquad\Rightarrow\qquad
\Xi^{xx} =- \Xi^{yy}.
\]

Using the explicit form of the collision integrals (\ref{iab0_1}) and
the quantities $\Xi^{i\beta}$, Eq.~(\ref{xivec}), and taking into
account the above argument, I can express the vector on the right-hand
side of Eq.~(\ref{schaeqB}) as
\[
-4
\frac{\alpha_g^2}{2\pi}\frac{NT^2}{\cal T} 
\widehat{\textswab{T}}_\eta
\widehat{\textswab{M}}_h^{-1}
\begin{pmatrix}
\delta\Pi^{\alpha\beta}/v_g^2 \cr
\delta\Pi^{\alpha\beta}_I /v_g^2\cr
\delta\Pi^{\alpha\beta}_E /T
\end{pmatrix}
-
8\omega_B
\widehat{\textswab{M}}_K
\widehat{\textswab{M}}_h^{-1}
\epsilon^{\alpha ji} e_B^j 
\begin{pmatrix}
\delta\Pi^{i\beta}/v_g^2 \cr
\delta\Pi^{i\beta}_I /v_g^2\cr
\delta\Pi^{i\beta}_E /T
\end{pmatrix}\!,
\qquad
\widehat{\textswab{T}}_\eta =
\widehat{\textswab{T}}_\Pi+\frac{1}{t_{\rm d}}\widehat{\textswab{M}}_h,
\]
where $\widehat{\textswab{T}}_\Pi$ is the matrix of electron-electron
scattering rates, see Eq.~(\ref{iab0_1}), in the dimensionless form
(\ref{mdim}). 

Now I can solve Eq.~(\ref{schaeqB}) similarly to the solution of
Eq.~(\ref{djeqsB}). Introducing the notation [cf. Eq.~(\ref{sxx})],
\begin{equation}
\label{hib}
\textgoth{h}^{\alpha\beta}
=
\widehat{\textswab{M}}_h^{-1}
\begin{pmatrix}
\delta\Pi^{i\beta}/v_g^2 \cr
\delta\Pi^{i\beta}_I /v_g^2\cr
\delta\Pi^{i\beta}_E /T
\end{pmatrix}\!,
\qquad
\textswab{D}^{\alpha\beta}=
\nabla^\alpha u^\beta+\nabla^\beta u^\alpha-\delta^{\alpha\beta}\bs{\nabla}\!\cdot\!\bs{u},
\qquad
\gamma_B = \frac{|e|v_g^2B}{\alpha_g^2cT^2},
\end{equation}
I may re-write Eq.~(\ref{schaeqB}) as
\[
\widehat{\textswab{T}}_\eta \textgoth{h}^{\alpha\beta}
+
\pi\gamma_B \widehat{\textswab{M}}_K\epsilon^{\alpha ji} e_B^j\textgoth{h}^{i\beta}
=
-
\frac{\cal T}{4\alpha_g^2v_g^2}
\begin{pmatrix}
\tilde{n} \cr
x^2/2\!+\!\pi^2/6 \cr
3\tilde{n}_{E}
\end{pmatrix}
\textswab{D}^{\alpha\beta}.
\]
Multiplying this equation by $\bs{e}_B$, I obtain [similarly to the
calculation below Eq.~(\ref{sxx})]
\[
-\pi\gamma_B \widehat{\textswab{M}}_K \textgoth{h}^{\alpha\beta}
+
\widehat{\textswab{T}}_\eta \epsilon^{\alpha ji} e_B^j\textgoth{h}^{i\beta}
=
-
\frac{\cal T}{4\alpha_g^2v_g^2}
\begin{pmatrix}
\tilde{n} \cr
x^2/2\!+\!\pi^2/6 \cr
3\tilde{n}_{E}
\end{pmatrix}
\epsilon^{\alpha ji} e_B^j \textswab{D}^{i\beta}.
\]
Similarly to Eq.~(\ref{sigmagen}), I find the solution in the form (I
am only interested in $\delta\Pi^{\alpha\beta}_E$)
\begin{subequations}
\label{etaB}
\begin{eqnarray}
\label{dpeb}
&&
\delta\Pi^{\alpha\beta}_E
=
- \frac{{\cal T}T}{4\alpha_g^2v_g^2}
\begin{pmatrix}
0 & 0 & 1
\end{pmatrix}
\widehat{\textswab{M}}_h
\left(1\!+\!\pi^2\gamma_B^2\widehat{\textswab{T}}_\eta^{-1}\widehat{\textswab{M}}_K
\widehat{\textswab{T}}_\eta^{-1}\widehat{\textswab{M}}_K
\right)^{\!\!-1}
\widehat{\textswab{T}}_\eta^{-1}
\begin{pmatrix}
\tilde{n} \cr
x^2/2\!+\!\pi^2/6 \cr
3\tilde{n}_{E}
\end{pmatrix}
\textswab{D}^{\alpha\beta}
\\
&&
\nonumber\\
&&
\qquad\qquad
+ \frac{{\cal T}T}{4\alpha_g^2v_g^2}
\begin{pmatrix}
0 & 0 & 1
\end{pmatrix}
\widehat{\textswab{M}}_h
\left(1\!+\!\pi^2\gamma_B^2\widehat{\textswab{T}}_\eta^{-1}\widehat{\textswab{M}}_K
\widehat{\textswab{T}}_\eta^{-1}\widehat{\textswab{M}}_K
\right)^{\!\!-1}
\widehat{\textswab{T}}_\eta^{-1}\widehat{\textswab{M}}_K\widehat{\textswab{T}}_\eta^{-1}
\begin{pmatrix}
\tilde{n} \cr
x^2/2\!+\!\pi^2/6 \cr
3\tilde{n}_{E}
\end{pmatrix}
\epsilon^{\alpha ji} e_B^j \textswab{D}^{i\beta}.
\nonumber
\end{eqnarray}
Generalizing the definition of the viscosity (\ref{dpi0}) to the case
of nonzero magnetic field,
\begin{equation}
\label{visdefB}
\delta\Pi^{\alpha\beta}_E=-\eta\textswab{D}^{\alpha\beta}
+\eta_H \epsilon^{\alpha ji} e_B^j \textswab{D}^{i\beta},
\end{equation}
\end{subequations}
I obtain the final expressions for the shear and Hall viscosities
\cite{ale,moo} in graphene
\begin{subequations}
\label{vr}
\begin{equation}
\label{etaB0}
\eta
=
\frac{{\cal T}T}{4\alpha_g^2v_g^2}
\begin{pmatrix}
0 & 0 & 1
\end{pmatrix}
\widehat{\textswab{M}}_h
\left(1+\pi^2\gamma_B^2\widehat{\textswab{T}}_\eta^{-1}\widehat{\textswab{M}}_K
\widehat{\textswab{T}}_\eta^{-1}\widehat{\textswab{M}}_K
\right)^{\!\!-1}
\widehat{\textswab{T}}_\eta^{-1}
\begin{pmatrix}
\tilde{n} \cr
x^2/2\!+\!\pi^2/6 \cr
3\tilde{n}_{E}
\end{pmatrix}\!,
\end{equation}
\begin{equation}
\label{etaH0}
\eta_H 
=\pi\gamma_B
\frac{{\cal T}T}{4\alpha_g^2v_g^2}
\begin{pmatrix}
0 & 0 & 1
\end{pmatrix}
\widehat{\textswab{M}}_h
\left(1+\pi^2\gamma_B^2\widehat{\textswab{T}}_\eta^{-1}\widehat{\textswab{M}}_K
\widehat{\textswab{T}}_\eta^{-1}\widehat{\textswab{M}}_K
\right)^{\!\!-1}
\widehat{\textswab{T}}_\eta^{-1}\widehat{\textswab{M}}_K\widehat{\textswab{T}}_\eta^{-1}
\begin{pmatrix}
\tilde{n} \cr
x^2/2\!+\!\pi^2/6 \cr
3\tilde{n}_{E}
\end{pmatrix}\!.
\end{equation}
\end{subequations}
The sign of the shear viscosity $\eta$ is fixed by the laws of
thermodynamics \cite{dau6,dau10}. In contrast, Hall viscosity is
non-dissipative (since the Lorentz force does not perform any work)
and may have an arbitrary sign which is technically determined by the
quasiparticle charge and direction of the magnetic field. In this
paper, I choose the $\eta_H$ to be positive for electrons by analogy
with Hall conductivity \cite{geim4}.

\subsection{Generalized Navier-Stokes equation}

Substituting the dissipative correction $\delta\Pi_E^{\alpha\beta}$
into the continuity equation for momentum density (\ref{cek}) and
repeating the steps used to derive the Euler equation (\ref{eq0g}) I
find the generalization of the central equation of the traditional
hydrodynamics, the Navier-Stokes equation \cite{dau6} to the
electronic system in graphene
\begin{equation}
\label{eq1g}
W(\partial_t+\bs{u}\!\cdot\!\bs{\nabla})\bs{u}
+
v_g^2 \bs{\nabla} P
+
\bs{u} \partial_t P 
+
e(\bs{E}\!\cdot\!\bs{j})\bs{u} 
=
v_g^2 
\left[
\eta \Delta\bs{u}
-
\eta_H \Delta\bs{u}\!\times\!\bs{e}_B
+
en\bs{E}
+
\frac{e}{c} \bs{j}\!\times\!\bs{B}
\right]
-
\frac{\bs{j}_E}{\tau_{{\rm dis}}}.
\end{equation}
Combined with the expressions for viscosities (\ref{vr}) and
dissipative corrections to currents (\ref{sigmagen}) this equation
represents the central result if this paper. Previously, the
generalized Navier-Stokes equation in graphene was derived in
Ref.~\cite{hydro1} in the absence of disorder and the magnetic field
and in Ref.~\cite{msf} in the absence of disorder and the external
fields (both electric and magnetic).

\section{Discussion}

The purpose of this paper was to derive the hydrodynamic equations for
the electronic fluid in graphene in the presence of electromagnetic
fields and weak disorder as well as to obtain closed expressions for
shear and Hall viscosities and electrical conductivity in graphene
(the latter being an analog of the thermal conductivity in the
traditional hydrodynamics). Despite the conceptual simplicity of the
assumptions leading to the hydrodynamic description, the Navier-Stokes
equation is known to yield a large number of important solutions, see
Ref.~\cite{dau6}. It is therefore impractical to include even a small
subset of these solutions into a single paper. Nevertheless it is
important to show that the cumbersome expressions for the dissipative
corrections to quasiparticle currents (\ref{sigmagen}) and the shear
viscosity (\ref{etaB0}) yield the well-known results in the simplest
limiting cases.

\subsection{Quantum conductivity}
\label{qc}

Consider first the electrical conductivity at charge neutrality in the
absence of the magnetic field, known as the ``quantum'' (or
``intrinsic'') conductivity \cite{luc,msf}. In this case the equation
(\ref{sigmagen}) simplifies. Setting the chemical potential to zero,
one also finds the vanishing charge density, ${\tilde{n}(0)=0}$, and
the ``imbalance compressibility'',
${\partial{n}_I/\partial\mu(0)=0}$. Hence the correction to the
electric current, $\delta\bs{j}$, represents the ``whole'' current,
since the ``ideal'' part of the current vanishes, see
Eq.~(\ref{hj0s}). At the same time, the Navier-Stokes equation
(\ref{eq1g}) [together with Eq.~(\ref{ljE})] admits a stationary and
uniform solution, ${\bs{u}=0}$. Therefore, for the stationary and
uniform fields the energy (as discussed above) and imbalance currents
vanish, see Ref.~\cite{hydro0},
\[
\bs{j}_I(\mu=0)=0, \qquad \bs{j}_E(\mu=0)=0.
\]
The combination of the distribution functions (\ref{iz0}) vanishes as
well, ${I(x=0)=0}$, such that ${\tau^{-1}_{12}(\mu\!=\!0)=0}$. The
energy density and compressibility are determined by temperature,
${n_E=3N\zeta(3)T^3/(2\pi{v}_g^2)}$, ${{\cal T}=2T\ln2}$. Substituting
these values into Eq.~(\ref{sigmagen}), I find
\begin{equation}
\label{jdisdp}
\begin{pmatrix}
      \delta\bs{j} \cr
      \delta\bs{j}_I \cr
      \delta\bs{j}_E
\end{pmatrix}\!
=
\frac{8\ln^22}{\alpha_g^2}\!
\begin{pmatrix}
1 & 0 & 0 \cr
0 & 1 & \frac{\pi^2}{6\ln2} \cr
0 & \frac{\pi^2}{6\ln2} & \frac{9\zeta(3)}{2\ln2}
\end{pmatrix}\!\!
\begin{pmatrix}
\frac{t_{11}t_{\rm d}}{t_{11}+t_{\rm d}} & 0 & 0 \cr
0 & \frac{t_{22}t_{\rm d}}{t_{22}\delta+t_{\rm d}} & 
-\frac{\pi^2}{27\zeta(3)} \frac{t_{22}t_{\rm d}}{t_{22}\delta+t_{\rm d}} \cr
0 & -\frac{\pi^2}{27\zeta(3)} \frac{t_{22}t_{\rm d}}{t_{22}\delta+t_{\rm d}} & 
\frac{2\ln2}{9\zeta(3)}\frac{(t_{22}+t_{\rm d})t_{\rm d}}{t_{22}\delta+t_{\rm d}}
\end{pmatrix}\!\!
\begin{pmatrix}
  1 & 0 & 0\cr
  0 & -\delta & 0\cr
  0 & 0 & 0
\end{pmatrix}\!\!
\begin{pmatrix}
  e\bs{E} \cr
  0 \cr
  0
\end{pmatrix}\!,
\end{equation}
where $t_{11}$ and $t_{22}$ are the diagonal elements of the matrix
$\widehat{\textswab{T}}$, see Eq.~(\ref{mdim}), and
\[
\delta = 1-\frac{\pi^4}{162\zeta(3)\ln2}.
\]

The solution (\ref{jdisdp}) for the dissipative correction to the
electric current yields the resistivity of undoped graphene
\cite{hydro0}
\begin{equation}
\label{sigma0}
R(\mu\!=\!0;\bs{B}\!=\!0)
=
\frac{\pi}{2e^2T\ln2}\left(\frac{1}{\tau_{11}}\!+\!\frac{1}{\tau_{\rm dis}}\right)
\underset{\tau_{\rm dis}\rightarrow\infty}{\longrightarrow} \frac{1}{\sigma_Q},
\qquad
\sigma_Q = \frac{8\ln^22}{\alpha_g^2} e^2 t_{11}(x=0) = {\cal A} \frac{e^2}{\alpha_g^2},
\end{equation}
where $\sigma_Q$ is the ``intrinsic'' conductivity of disorder-free
graphene at the Dirac point \cite{das,kash,mfss,mfs,schutt}. Previously,
the coefficient ${\cal A}$ was reported to have values ${{\cal
    A}=0.12}$ \cite{kash} and ${{\cal{A}}=0.19}$
\cite{hydro1}. Evaluating the integral for $t_{11}(0)$ numerically for
unscreened Coulomb interaction (as was done in
Refs.~\cite{kash,hydro1}), I obtain the value
${{\cal{A}}=0.113\pm0.03}$, where the error comes from systematic
differences between various numerical methods (note, that the present
calculation neglects the exchange contribution $\sim1/N$, which was
shown to be numerically small in Ref.~\cite{kash}).

\subsection{Electrical conductivity in the degenerate regime}

Away from charge neutrality, the electric current is no longer
determined by the dissipative correction alone. The ``ideal''
contribution to the current is governed by the hydrodynamic velocity
$\bs{u}$, see Eq.~(\ref{hj0s}). The velocity $\bs{u}$ is a solution of
the Navier-Stokes equation (\ref{eq1g}). Within linear response and in
the absence of the magnetic field, the uniform and time-independent
solution to Eq.~(\ref{eq1g}) is simply the Ohm's law with the Drude
formula for conductivity
\[
v_g^2 en\bs{E}
-
\frac{W\bs{u}}{\tau_{{\rm dis}}}
=0
\qquad\Rightarrow\qquad
\bs{u}=\frac{ev_g^2\tau_{{\rm dis}}}{\mu}\bs{E}.
\]
In the absence of magnetic field, Eq.~(\ref{ljE}) yields the vanishing
dissipative correction to the energy current,
$\delta\bs{j}_E=0$. However, in the degenerate regime, ${\mu\gg{T}}$,
all three dissipative corrections $\delta\bs{j}$, $\delta\bs{j}_I$,
and $\delta\bs{j}_E$ are proportional to each other, since the three
rows of the matrix $\widehat{M}_h$ [see Eq.~(\ref{djs})] are identical
(in this limit ${n=n_I=\mu^2/(\pi{v}_g^2)}$ since only one band
contributes and ${n_E=2\mu^3/(3\pi{v}_g^2)}$, with $N=4$). This means
that all three dissipative corrections vanish and, in particular,
\[
\delta\bs{j}(\mu\gg T)=0.
\]
The total electric current $\bs{J}=e\bs{j}$ is then determined by the
``hydrodynamic'' contribution alone 
\begin{equation}
\label{sdrude}
\bs{J}=en\bs{u}
=
e^2\frac{1}{\pi}\mu\tau_{{\rm dis}} \bs{E},
\qquad\Rightarrow\qquad
\sigma(\mu\gg T) \approx \frac{e^2}{\pi}\mu\tau_{{\rm dis}}.
\end{equation}
Thus, the conductivity of graphene in degenerate regime is given by
the Drude result (due to disorder), in agreement with the
leading-order result of the linear response theory \cite{hydro0}.

\subsection{Magnetoconductivity at charge neutrality}

Similarly to the discussion in Sec.~\ref{qc}, the equation for the
dissipative corrections (\ref{djeqsB}) simplifies at charge
neutrality. Instead of using the general solution (\ref{sigmagen}), it
might be instructive to represent Eq.~(\ref{djeqsB}) as three vector
equations and solve them directly. Indeed, using the parameter values
listed in Sec.~\ref{qc} together with the stationary and uniform
solution of the Navier-Stokes equation, ${\bs{u}=0}$, one can
represent the matrix equation (\ref{djeqsB}) as a system of coupled
equations
\begin{eqnarray*}
&&
e\bs{E} = \frac{\alpha_g^2}{8\ln^22}\left[\frac{1}{t_{11}}\!+\!\frac{1}{t_{\rm d}}\right]\delta\bs{j}
+
\frac{\pi\omega_B}{2T\delta\ln2}
\left[1\!-\!\frac{2\pi^2\ln2}{27\zeta(3)}\right]\bs{e}_B\!\times\!\delta\bs{j}_I
-
\frac{\pi\omega_B}{18T^2\delta\zeta(3)\ln2}
\left[\frac{\pi^2}{3}\!-\!4\ln^22\right]\bs{e}_B\!\times\!\delta\bs{j}_E,
\\
&&
\\
&&
0 =  \frac{\alpha_g^2}{8\ln^22}\left[\frac{1}{t_{22}\delta}\!+\!\frac{1}{t_{\rm d}}\right]\delta\bs{j}_I
-
\frac{\alpha_g^2}{8\ln^22}\frac{1}{t_{22}T}\frac{\pi^2}{27\delta\zeta(3)}\delta\bs{j}_E
+ 
\frac{\pi\omega_B}{2T\ln2} \bs{e}_B\!\times\!\delta\bs{j},
\\
&&
\\
&&
0 =\frac{\alpha_g^2}{8\ln^22}\frac{1}{t_{\rm d}T}\delta\bs{j}_E
+
\frac{\pi\omega_B}{T} \bs{e}_B\!\times\!\delta\bs{j}.
\end{eqnarray*}
The last equation yields the energy current
\begin{subequations}
\label{dpsigres}
\begin{equation}
\label{dpesig}
\delta\bs{j}_E = - (2\ln2) \omega_B T\tau_{\rm dis} \bs{e}_B\!\times\!\delta\bs{j}.
\end{equation}
Substituting that result into the preceding equation, one finds for the imbalance current
\begin{equation}
\label{dpisig}
\delta\bs{j}_I = - 
\frac{1\!+\!\frac{2\pi^2\ln2}{27\zeta(3)\delta}\frac{\tau_{\rm dis}}{\tau_{22}}}
{1\!+\!\frac{1}{\delta}\frac{\tau_{\rm dis}}{\tau_{22}}}
\omega_B\tau_{\rm dis} \bs{e}_B\!\times\!\delta\bs{j}.
\end{equation}
Finally, excluding the imbalance and energy currents from the first equation, one finds
\begin{equation}
\label{dpjsig}
\bs{E} = R(\mu\!=\!0;\bs{B}\!=\!0)\delta\bs{J}
+
\frac{\pi\omega^2_B\tau_{\rm dis}}{e^2T\delta}
\frac{\frac{\pi^2}{3}\!-\!4\ln^22}{9\zeta(3)}
\left[
-1\!+\!\frac{9\zeta(3)\!-\!\frac{2\pi^2}{3}\ln2}{2\ln2\left(\frac{\pi^2}{3}\!-\!4\ln^22\right)}
\frac{1\!+\!\frac{2\pi^2\ln2}{27\zeta(3)\delta}\frac{\tau_{\rm dis}}{\tau_{22}}}
{1\!+\!\frac{1}{\delta}\frac{\tau_{\rm dis}}{\tau_{22}}}
\right]
\delta\bs{J}.
\end{equation}
\end{subequations}
The fact that the electric current in magnetic field is parallel to
the electric field can be expressed in terms of vanishing Hall
coefficient, physically due to the exact electron-hole symmetry,
\begin{subequations}
\label{rBres}
\begin{equation}
\label{rh0}
R_H(\mu\!=\!0)=0.
\end{equation}
At the same time, Eq.~(\ref{dpjsig}) yields {\it positive}, longitudinal
magnetoresistance (previously found in Refs.~\cite{hydro0,mus})
\begin{equation}
\label{rB0}
R(B;\mu\!=\!0) = R(B\!=\!0;\mu\!=\!0) + \delta R(B;\mu\!=\!0),
\qquad
\delta R(B;\mu\!=\!0) = {\cal C} \frac{v_g^4}{c^2}\frac{B^2\tau_{\rm dis}}{T^3},
\end{equation}
where
\begin{equation}
\label{rB0c}
{\cal C} = \frac{\pi}{9\zeta(3)}\frac{\frac{\pi^2}{3}\!-\!4\ln^22}{4\delta\ln^22}
\left[
\frac{9\zeta(3)\!-\!\frac{2\pi^2}{3}\ln2}{2\ln2\left(\frac{\pi^2}{3}\!-\!4\ln^22\right)}
\frac{1\!+\!\frac{2\pi^2\ln2}{27\zeta(3)\delta}\frac{\tau_{\rm dis}}{\tau_{22}}}
{1\!+\!\frac{1}{\delta}\frac{\tau_{\rm dis}}{\tau_{22}}}
\!-\!1
\right]
\approx
\frac{1.71\!+\!1.04\tau_{\rm dis}/\tau_{22}}{1\!+\!3.59\tau_{\rm dis}/\tau_{22}}
\underset{\tau_{\rm dis}\rightarrow\infty}{\longrightarrow}
\frac{\pi}{9\zeta(3)}\approx0.2904.
\end{equation}
\end{subequations}

In this section I have assumed an infinite system. Hence all
macroscopic quantities are homogeneous and the hydrodynamic equations
become algebraic. In finite size systems the situation is more
involved. Here one has to solve differential equations (with the
appropriate boundary conditions \cite{ks19}). The resulting flows may
be highly inhomogeneous \cite{fl0,cfl,mr1,msw2,msw} leading to
e.g. linear \cite{hydro0,mr1,mrexp} or negative magnetoresisteance
\cite{ale,lucnmr,mr2}.

\subsection{Shear viscosity at charge neutrality in zero field}

The expression for the shear viscosity also simplifies at charge
neutrality (the Hall viscosity vanishes at this point altogether). The
vanishing combination of the distribution functions (\ref{iz0}),
${I(x=0)=0}$, leads to vanishing of the two scattering rates
\[
\widetilde{Y}_{01}(\mu=0)=\widetilde{Y}_{12}(\mu=0)=\widetilde{Y}_{13}(\mu=0)=0
\qquad\Rightarrow\qquad
\tilde\tau_{12}^{-1}(\mu=0)=\tilde\tau_{13}^{-1}(\mu=0)=0.
\]
In the absence of the magnetic field, one can simplify the
coefficients in Eq.~(\ref{etaB0}) using the parameter values listed in
Sec.~\ref{qc}. This leads to the following expression for the shear
viscosity
\begin{eqnarray*}
&&
\eta = \frac{T^2}{4\alpha_g^2v_g^2}
\begin{pmatrix}
 0 & 0 & 1
\end{pmatrix}
\begin{pmatrix}
2\ln2 & 0 & 0 \cr
0 & 2\ln2 & \frac{\pi^2}{3} \cr
0 & \frac{\pi^2}{3} & 9\zeta(3)  \cr
\end{pmatrix}
\begin{pmatrix}
\tilde{t}_{11}^{-1} & 0 & 0 \cr
0 & \tilde{t}_{22}^{-1} & \tilde{t}_{23}^{-1} \cr
0 & \tilde{t}_{32}^{-1} & \tilde{t}_{33}^{-1}
\end{pmatrix}^{-1}
\begin{pmatrix}
0 \cr
\pi^2/6 \cr
9\zeta(3)/2
\end{pmatrix}
\\
&&
\\
&&
\qquad
=
\frac{T^2}{8\alpha_g^2v_g^2}
\frac{81 \zeta(3)^2 \tilde{t}_{23}\tilde{t}_{33}+(\pi^4/9)\tilde{t}_{22}\tilde{t}_{23}
- 6\pi^2\zeta(3) \tilde{t}_{22}\tilde{t}_{33}}{\tilde{t}_{23}^2-\tilde{t}_{22}\tilde{t}_{33}}
\tilde{t}_{23},
\end{eqnarray*}
and hence
\begin{equation}
\label{vdp}
\eta(\mu=0) = {\cal B} \frac{T^2}{\alpha_g^2 v_g^2},
\qquad
{\cal B} = 
\frac{\tilde{t}_{23}}{8}
\frac{81 \zeta(3)^2 \tilde{t}_{23}\tilde{t}_{33}+(\pi^4/9)\tilde{t}_{22}\tilde{t}_{23}
- 6\pi^2\zeta(3) \tilde{t}_{22}\tilde{t}_{33}}{\tilde{t}_{23}^2-\tilde{t}_{22}\tilde{t}_{33}},
\end{equation}
where $\tilde{t}_{ij}$ are the elements of the matrix
$\widehat{\textswab{T}}_\eta$. This result was previously found in
Ref.~\cite{msf} with the numerical value of the coefficient
${{\cal{B}}=0.45}$ (evaluated for unscreened Coulomb interaction in
the absence of disorder). Evaluating the dimensionless scattering
rates $t_{ij}$ numerically, I find ${{\cal{B}}=0.446\pm0.005}$ where
similarly to the coefficient ${\cal A}$ in Eq.~(\ref{sigma0}) the
deviation is due to differences between various numerical methods.
The exchange contribution is again neglected, but this does not seem
to lead to any appreciable error.

Finally, I can use Eq.~(\ref{eng}) to compute the ratio of the shear
viscosity to the entropy density at charge neutrality. Indeed, for
${\mu_\pm=0}$ the entropy density is determined by pressure,
${s=P/T}$, which in turn is proportional to the energy density. As a result,
\begin{equation}
\label{ent0}
s(\mu_\pm=0; \bs{u}=0) \approx \frac{9N\zeta(3)T^2}{4\pi v_g^2}.
\end{equation}
Dividing the viscosity (\ref{vdp}) by the entropy density (\ref{ent0})
one finds
\begin{equation}
\frac{\eta(\mu=0)}{s(\mu=0)} = {\cal B} \frac{T^2}{\alpha_g^2 v_g^2}
\frac{4\pi}{9N\zeta(3)}\frac{v_g^2}{T^2} = \frac{\pi{\cal
    B}}{9\zeta(3)} \frac{1}{\alpha_g^2} = \frac{0.131}{\alpha_g^2}.
\end{equation}
This should be compared with the conjectured lower bound \cite{kov}
\begin{equation}
\frac{\eta}{s} \geqslant \frac{1}{4\pi}\approx 0.0796.
\end{equation}
The shear viscosity to entropy ratio was discussed in detail in
\cite{msf}, where it was shown that renormalization of the coupling 
constant leads to a logarithmic temperature dependence of the above
ratio [formally, by replacing $\alpha^2$ in denominator by the
renormalized value $\alpha^2(T)\approx16/\ln^2(T_\Lambda/T)$, where
$T_\Lambda$ is the cut-off scale]. Hence the ratio is expected to grow
as one lowers the temperature. At high enough temperatures, the ratio
decreases with the growing $T$, never quite reaching the lower
bound \cite{msf}.

\section{Summary}

In this paper I have presented a detailed derivation of the
hydrodynamic theory of electronic transport in graphene in the
presence of the external magnetic field and weak disorder. The main
results of the paper are the generalized Navier-Stokes equation
(\ref{eq1g}), explicit expressions for the shear and Hall viscosity
(\ref{vr}) and for the dissipative corrections to the quasiparticle
currents (\ref{sigmagen}). These results agree with the previously
reported values of the quantum conductivity
\cite{kash,mfss,mfs,schutt} and shear viscosity \cite{msf} at charge
neutrality in pure graphene in zero field, providing an extension to
arbitrary doping levels, (non-quantizing) magnetic fields, and weak
disorder. For a detailed discussion of the quantitative results for
the shear and Hall viscosities in a wide range of temperatures and
carrier densities and their relation to the experimental data of
Refs.~\cite{geim4,geim1} see Ref.~\cite{me2}. For a similar discussion
of the optical conductivity see Ref.~\cite{me3}.

The hydrodynamic theory derived in this paper is justified by the
inequality (\ref{tc}) and is valid for classical (non-quantizing)
magnetic fields. The former condition restricts the temperature range
of the hydrodynamic effects to relatively high temperatures above
$100$K (for a discussion of the experimental viewpoint on that issue
see Ref.~\cite{geim3}). The latter requirement was discussed in detail
in Ref.~\cite{mfs} (see also the corresponding discussion in
Ref.~\cite{me3}). At high enough temperatures [required by
  Eq.~(\ref{tc})] Landau levels are smeared out such that one does not
need to account for the quantum Hall effect for magnetic fields up to
several Tesla. At stronger fields the hydrodynamic description breaks
down. However, the kinetic theory has a wider applicability range. As
a result, magnetotransport in stronger fields can be described by the
standard linear response theory \cite{mfs,hydro0,me3}. Such
considerations are applicable as long as the typical cyclotron
frequency does not exceed temperature \cite{mfs}. Beyond that field
range, quantum effects take over which cannot be described by the
semiclassical kinetic theory. Transport in quantized magnetic fields
is beyond the scope of the present paper.

Very recently I became aware of the related work on magnetotransport in
graphene, Ref.~\cite{ady2}, and on Hall viscosity, Refs.~\cite{bur19,ady}.

\section*{Acknowledgments}

The author wishes to thank I.V. Gornyi, A.D. Mirlin, J. Schmalian,
M. Sch\"utt, and A. Shnirman for fruitful discussions. This work was
supported by the German Research Foundation DFG within FLAG-ERA Joint
Transnational Call (Project GRANSPORT), by the European Commission
under the EU Horizon 2020 MSCA-RISE-2019 program (Project 873028
HYDROTRONICS), and the MEPhI Academic Excellence Project, Contract
No. 02.a03.21.0005.

%%%%%%%%%%%%%%%%%%%%%%%%%%%%%%%%%%%%%%%%%%%%%%%%%%%%%%%%%%%%%%%%%%%%%%%%%
%%%%%%%%%%%%%%%%%%%%%%%%%%%%%%%%%%%%%%%%%%%%%%%%%%%%%%%%%%%%%%%%%%%%%%%%%

\appendix

\section{Macroscopic quantities}
\label{appA}

\subsection{Quasiparticle densities}
\label{qds}

Quasiparticle (number) density can be defined in the usual form
\cite{dau10}. The only subtle point arising in two-band systems is the
treatment of the formally infinite number of particles in the filled
band. In other words, one needs to make a distinction between the
``particle density'' and ``carrier density''. In the valence band, the
latter is defined by the particle-hole transformation
\[
1 \!-\! f_{-,\bs{k}} \!=\! f_{-,\bs{k}}(-[\epsilon_{-,\bs{k}}\!-\!\mu_-])
\!\equiv\! f_{h,\bs{k}}(\epsilon_{h,\bs{k}}\!-\!\mu_h),
\]
with the hole energies and chemical potential defined as
\begin{equation}
\label{eh}
\epsilon_{h,\bs{k}} \equiv - \epsilon_{-,\bs{k}} = v_gk,
\qquad
\mu_h \equiv -\mu_-.
\end{equation}

The numbers of ``carriers'' (i.e. the low-energy excitations) in the
two bands can be explicitly defined as
\begin{subequations}
\label{den}
\begin{equation}
\label{np}
n_+=N\!\int\!\!\frac{d^2k}{(2\pi)^2} f_{+,\bs{k}},
\end{equation} 
and (without introducing the hole index)
\begin{equation}
\label{nm}
n_-=N\!\int\!\!\frac{d^2k}{(2\pi)^2} \left(1-f_{-,\bs{k}}\right),
\end{equation}
with the total ``charge'' (or ``carrier'') density being
\begin{equation}
\label{n}
n = n_+ - n_-.
\end{equation}
Summing up the densities (\ref{np}) and (\ref{nm}), one may
define the ``imbalance'' or the total quasiparticle density
\begin{equation}
\label{ni}
n_I = n_+ + n_-.
\end{equation}
\end{subequations}

\subsection{Quasiparticle currents}
\label{qpcs}

Similarly to the quasiparticle densities (\ref{den}), the macroscopic
currents can be defined as
\begin{subequations}
\label{js}
\begin{equation}
\label{jp}
\bs{j}_+ =N\!\int\!\!\frac{d^2k}{(2\pi)^2} \bs{v}_{+,\bs{k}} f_{+,\bs{k}},
\end{equation}
\begin{equation}
\label{jm}
\bs{j}_- =N\!\int\!\!\frac{d^2k}{(2\pi)^2} \bs{v}_{-,\bs{k}} \left(1\!-\!f_{-,\bs{k}}\right),
\end{equation}
\begin{equation}
\label{j}
\bs{j} \!=\! \bs{j}_+ \!-\! \bs{j}_- \!=\! 
N\!\!\int\!\!\frac{d^2k}{(2\pi)^2} 
\left[\bs{v}_{+,\bs{k}} f_{+,\bs{k}} \!-\! \bs{v}_{-,\bs{k}}\left(1\!-\!f_{-,\bs{k}}\right)\right],
\end{equation}
\begin{equation}
\label{ji}
\bs{j}_I = \bs{j}_+ + \bs{j}_- .
\end{equation}
\end{subequations}

\subsection{Energy density}
\label{neapp}

In two-band systems with unbound (from below) spectrum, one has to
define the energy density relative to the (formally infinite) energy
of the filled valence band
\begin{subequations}
\label{ne}
\begin{equation}
n_E = N\!\sum_\lambda\!\int\!\!\frac{d^2k}{(2\pi)^2} \epsilon_{\lambda\bs{k}}
f_{\lambda\bs{k}} - n_{E}^{(0)},
\end{equation}
\begin{equation}
\label{neinf}
n_{E}^{(0)} = N\! \!\int\!\!\frac{d^2k}{(2\pi)^2} \epsilon_{-,\bs{k}}.
\end{equation}
This is equivalent to the electron-hole transformation (\ref{eh})
based on the following observation
\begin{equation}
\label{ne1}
n_E = N\!\int\!\!\frac{d^2k}{(2\pi)^2} 
\left[ \epsilon_{+,\bs{k}} f_{+,\bs{k}} + \epsilon_{-,\bs{k}} \left(f_{-,\bs{k}}\!-\!1\right)\right].
\end{equation}
As a result, one may re-write Eq.~(\ref{ne1}) as
\begin{equation}
\label{ne2}
n_E = N\!\int\!\!\frac{d^2k}{(2\pi)^2} 
\left[ \epsilon_{e,\bs{k}} f_{e,\bs{k}} + \epsilon_{h,\bs{k}} f_{h,\bs{k}}\right],
\end{equation}
\end{subequations}
where the subscript ``e'' stands for ``electrons'' and replaces the
index $+$ in Eq.~(\ref{ne1}), while the subscript ``h'' stands for
``holes'' as defined in Eq.~(\ref{eh}).

\subsection{Energy current}
\label{jeapp}

The calculation of the energy current can be performed along the same
lines. In graphene, the energy current, $\bs{j}_E$, is defined as
\begin{subequations}
\label{je}
\begin{equation}
\label{je1}
\bs{j}_E = N\!\sum_\lambda\!\int\!\frac{d^2k}{(2\pi)^2} 
\epsilon_{\lambda\bs{k}} \bs{v}_{\lambda\bs{k}} f_{\lambda\bs{k}}.
\end{equation}
In terms of band contributions, the energy current has the form
\begin{equation}
\label{je11}
\bs{j}_E = \bs{j}_{E+} + \bs{j}_{E-},
\end{equation}
where
\begin{equation}
\bs{j}_{E+} = N\!\int\!\frac{d^2k}{(2\pi)^2} 
\epsilon_{+\bs{k}} \bs{v}_{+\bs{k}} f_{+\bs{k}},
\end{equation}
\begin{equation}
\label{je11m}
\bs{j}_{E-} = N\!\int\!\frac{d^2k}{(2\pi)^2} 
\epsilon_{-\bs{k}} \bs{v}_{-\bs{k}} \left(f_{-\bs{k}}-1\right).
\end{equation}
The additional unity in (\ref{je11m}) serves to demonstrate
convergence, although the integral with unity vanishes anyway due to
rotation invariance.

Alternatively, one may re-write the energy current (\ref{je1}) using
the electron-hole transformation (\ref{eh}) as
\begin{equation}
\label{je2}
\bs{j}_E \!=\! \bs{j}_{E,e} \!+\! \bs{j}_{E,h}, \qquad
\bs{j}_{E,h} \!=\! 
- N \!\!\int\!\!\frac{d^2k}{(2\pi)^2} \epsilon_{h,\bs{k}}\bs{v}_{h,\bs{k}}  f_{h,\bs{k}}.
\end{equation}
\end{subequations}
Note, that in graphene the energy current (\ref{je}) is proportional
to the momentum density [due to Eq.~(\ref{vg})]
\begin{equation}
\label{nk}
\bs{n}_{\bs{k}} = N\!\sum_\lambda\!\int\!\frac{d^2k}{(2\pi)^2} \bs{k} f_{\lambda\bs{k}}
=
v_g^{-2}\bs{j}_E.
\end{equation}

\subsection{Momentum flux tensor}
\label{pab0}

Similar calculation can be performed for the momentum flux tensor
(also known as the stress-energy tensor or the energy-momentum tensor)
\begin{equation}
\label{pab}
\Pi_E^{\alpha\beta} = N\!\sum_\lambda\!\int\!\!\frac{d^2k}{(2\pi)^2} 
k^\alpha v^\beta_{\lambda\bs{k}} f_{\lambda\bs{k}}.
\end{equation}
Formally, the expression (\ref{pab}) is divergent and (similarly to
the quasiparticle and energy densities) has to be defined up to the
formally in finite contribution of the filled band. However, in all
hydrodynamic equations I will be dealing with derivatives of
$\Pi_E^{\alpha\beta}$ which allows me to subtract this
contribution. Note, that in a rotationally invariant system, the
kinetic definition (\ref{pab}) is manifestly symmetric.

In addition two further tensor quantities can be formed (by analogy
with the three macroscopic currents): the ``velocity flux tensor''
\begin{equation}
\label{piab}
\Pi^{\alpha\beta}
=
N\sum_\lambda\int\!\frac{d^2k}{(2\pi)^2} 
v^\alpha_{\lambda\bs{k}}v^\beta_{\lambda\bs{k}} f_{\lambda\bs{k}}
=
\sum_\lambda \Pi^{\alpha\beta}_{\lambda},
\end{equation}
and the ``imbalance flux tensor''
\begin{equation}
\label{piIab}
\Pi^{\alpha\beta}_{I}
=
N\sum_\lambda\int\!\frac{d^2k}{(2\pi)^2} \lambda
v^\alpha_{\lambda\bs{k}}v^\beta_{\lambda\bs{k}} f_{\lambda\bs{k}}
=
\sum_\lambda \lambda \Pi^{\alpha\beta}_{\lambda},
\end{equation}
cf. Eqs.~(\ref{pab}) and (\ref{piab}). These quantities are not
directly related to any traditional observables and play an auxiliary
role in the kinetic theory.

\subsection{Pressure and enthalpy}
\label{pwapp}

Under the assumption of local equilibrium, the expression for pressure
can be obtained from the thermodynamic potential since \cite{dau10}
\begin{equation}
\label{tp}
\Omega = - PV.
\end{equation}
The thermodynamic potential is calculated in the usual grand canonical
ensemble, with the only caveat that one should be careful while
dealing with the nearly filled valence band. The easiest way is to use
the electron-hole transformation (\ref{eh}). Then the thermodynamic
potential of the two-band system described by the local equilibrium
distribution (\ref{le}) can be written as
\begin{equation}
\label{omega}
\Omega = - T V N\!\int\!\frac{d^2k}{(2\pi)^2} 
\ln \left[1 + e^{(\mu_+\!-\epsilon_{+,\bs{k}}\!+\bs{u}\cdot\bs{k})/T}\right]
- T V N\!\int\!\frac{d^2k}{(2\pi)^2} 
\ln \left[1 + e^{(\epsilon_{-,\bs{k}}\!-\mu_-\!-\bs{u}\cdot\bs{k})/T}\right].
\end{equation}
Differentiating $\Omega$ with respect to the
chemical potential, one recovers the number of particles (\ref{den}):
\[
-\frac{\partial\Omega}{\partial\mu_+} =
V n_{+},
\qquad
\frac{\partial\Omega}{\partial\mu_-} =
V n_{-}
\qquad\Rightarrow\qquad
-\frac{\partial\Omega}{\partial\mu_+} - \frac{\partial\Omega}{\partial\mu_-} =
Vn.
\]
Combining Eqs.~(\ref{tp}) and (\ref{omega}), one finds the hydrodynamic pressure,
\begin{equation}
\label{p}
P = T N\!\int\!\frac{d^2k}{(2\pi)^2} 
\ln \left[1 + e^{(\mu_+\!-\epsilon_{+,\bs{k}}\!+\bs{u}\cdot\bs{k})/T}\right]
+
T N\!\int\!\frac{d^2k}{(2\pi)^2} 
\ln \left[1 + e^{(\epsilon_{-,\bs{k}}\!-\mu_-\!-\bs{u}\cdot\bs{k})/T}\right].
\end{equation}

The enthalpy can then be found using the standard relation 
\begin{equation}
\label{enthapp}
W=n_{E}+P.
\end{equation}

\subsection{Entropy}
\label{sapp}

In thermodynamics, entropy is defined as a 
\begin{equation}
\label{sdef}
S = - \left(\frac{\partial\Omega}{\partial T}\right)_{\mu},
\end{equation}
where $\Omega$ is the thermodynamic potential (\ref{omega}). A
straightforward calculation leads to the result:
\begin{equation}
\label{sd1}
TS = - \Omega + V n_{E,0 }-V(\mu_+n_{+,0} - \mu_-n_{-,0}) - V \bs{u}\cdot\bs{n}_{\bs{k}}.
\end{equation}
This should be compared with the standard thermodynamic relation
\[
E = \mu N + TS - PV =  \mu N + TS + \Omega.
\]
Given that the standard definition of entropy is formulated in thermal
equilibrium, it is not surprising that the two relations coincide for
$\bs{u}=0$ up to one important issue. Assuming the two independent
chemical potentials in graphene, one has to generalize Eq.~(\ref{sd1})
replacing
\[
\mu N \rightarrow \mu_+ N_+ - \mu_-N_-.
\]
and adding the velocity term. Then the entropy in local equilibrium is
given by
\begin{subequations}
\label{eng}
\begin{equation}
S = \left(E-\Omega-\mu_+ N_+ + \mu_-N_- - V \bs{u}\cdot\bs{n}_{\bs{k}}\right)/T.
\end{equation}
Dividing this expression by the volume and substituting the explicit
expressions, I find (for $\bs{u}\ne0$)
\begin{equation}
s = \left(3P -\mu_+ n_+ + \mu_-n_-\right)/T.
\end{equation}
\end{subequations}

\section{Nonequilibrium distribution function}
\label{ndflr}

Here I compare the dissipative correction to the local equilibrium
distribution function in the hydrodynamic approach to the
nonequilibrium distribution function considered within the linear
response theory of Ref.~\cite{hydro0}.

Within linear response, one starts with the global equilibrium
[described by the usual Fermi distribution function,
$f_0(\epsilon_{\lambda\bs{k}})$] where no currents are flowing
\[
N\!\sum_\lambda\!\!\int\!\!\frac{d^2k}{(2\pi)^2} \bs{v}_{\lambda\bs{k}} f_0(\epsilon_{\lambda\bs{k}})
=
N\!\sum_\lambda\lambda\!\!\int\!\!\frac{d^2k}{(2\pi)^2} \bs{v}_{\lambda\bs{k}} f_0(\epsilon_{\lambda\bs{k}})
=
N\!\sum_\lambda\!\int\!\!\frac{d^2k}{(2\pi)^2} 
\epsilon_{\lambda\bs{k}}\bs{v}_{\lambda\bs{k}} f_0(\epsilon_{\lambda\bs{k}})
=0.
\]
When the system is subjected to external fields, the distribution
function acquires a nonequilibrium correction, $\delta f_{LR}$,
yielding non-zero currents
\begin{equation}
\label{lrcs}
\bs{j} = N\!\sum_\lambda\!\!\int\!\!\frac{d^2k}{(2\pi)^2} \bs{v}_{\lambda\bs{k}} \delta f_{LR},
\qquad
\bs{j}_I = N\!\sum_\lambda\lambda\!\!\int\!\!\frac{d^2k}{(2\pi)^2} \bs{v}_{\lambda\bs{k}} \delta f_{LR},
\qquad
\bs{j}_E = N\!\sum_\lambda\lambda\!\int\!\!\frac{d^2k}{(2\pi)^2} 
\epsilon_{\lambda\bs{k}}\bs{v}_{\lambda\bs{k}} \delta f_{LR}.
\end{equation}
In the notation used in Ref.~\cite{hydro0}, ${\bs{P}\equiv\bs{j}_I}$, ${\bs{Q}\equiv\bs{j}_E}$.

The nonequilibrium correction, $\delta f_{LR}$, can be analyzed within
the same three mode approximation (\ref{hs}). Comparing the notation
used in the present paper to that of Ref.~\cite{hydro0}, one finds
[see Eq.~(\ref{df})]
\begin{equation}
\label{lrh}
h_{LR} = \frac{\bs{v}_{\lambda\bs{k}}}{v_g}\sum_{i=1}^3 \phi_i \bs{h}_{LR}^{(i)}
\equiv
\frac{2\bs{v}_{\lambda\bs{k}}}{\nu_0 {\cal T} v_g^2}
\left[\bs{\cal A}+\bs{\cal B}\frac{\epsilon}{\cal T}+\bs{\cal C}\lambda\right].
\end{equation}
Substituting the latter expression into the definitions (\ref{lrcs}),
one can express the quantities $\bs{\cal A}$, $\bs{\cal B}$, and
$\bs{\cal C}$ in terms of the macroscopic currents $\bs{j}$,
$\bs{j}_I$,a nd $\bs{j}_E$. In Ref.~\cite{hydro0} this was done
explicitly, but the result coincides with Eq.~(\ref{djs}). The reason
for this is that within the three-mode approximation the corrections
(\ref{disjs}) and the currents (\ref{lrcs}) are given by exact same
integrals. 

To better understand the relation between the two different
approaches, one can expand the local equilibrium distribution function
(\ref{le}) to the linear order in the hydrodynamic velocity. As a
result, the complete distribution function out of equilibrium takes
the form
\begin{equation}
\label{dfcom}
f = f^{(0)}+\delta f = f_0 + \delta f_{\bs{u}} + \delta f \rightarrow
f_0 + \delta f_{LR}.
\end{equation}
Representing $\delta f_{\bs{u}}$ in the three-mode form (\ref{hs})
\[
\bs{h}_{\bs{u}}^{(1)}=\bs{h}_{\bs{u}}^{(2)}=0, 
\qquad
\bs{h}_{\bs{u}}^{(3)}=\frac{\bs{u}}{v_g},
\]
and using Eq.~(\ref{djs}), I recover the ``ideal'' values for the
three macroscopic currents, Eqs.~(\ref{hj0s}) and (\ref{hje}), where
all nonlinearities (e.g., in denominators) are neglected.
Consequently, the sum of the two corrections
${\delta{f}_{\bs{u}}\!+\!\delta f}$ yields the total currents and
hence is equivalent to the linear response correction $\delta f_{LR}$.

\section{Integrated collision integral}

\subsection{Collision integral in the current equations}
\label{ciceqs}

Here I evaluate the collision integrals in the integrated kinetic
equations. Given the additive nature of the collision integral in the
kinetic equation (\ref{ke}), I separate the momentum-conserving
collision integral due to electron-electron interaction from disorder
scattering and other momentum nonconserving processes
\begin{equation}
\label{colintsep}
\bs{\cal I}_i\left[f\right] = \bs{\cal I}^{ee}_i\left[\delta f\right]
+ \bs{\cal I}^{\rm dis}_i\left[f\right],
\end{equation}
where the collision integral due to electron-electron interaction is
nullified by the local equilibrium distribution function,
${\bs{\cal{I}}^{ee}_i\left[f^{(0)}\right]=0}$, and all momentum non
conserving processes are grouped together into ${\bs{\cal I}^{\rm
    dis}_i\left[f\right]}$. These will be considered within the
simplest $\tau$-approximation.

\subsubsection{Collision integral due to electron-electron interaction}

The general form of the collision integral due to electron-electron
interaction is given by Eq.~(\ref{collint}). Introducing the
transferred energy $\omega$ and momentum $\bs{q}$, one may write the
transition probability (\ref{w0}) as
\begin{equation}
\label{w}
W_{12,1'2'} = \!\!\int\!\!\frac{d^2q}{(2\pi)^2}\frac{d\omega}{2\pi}
|U(\omega,\bs{q})|^2\left|\lambda_{\bs{v}_{1}\bs{v}_{1'}}\right|^2
\left|\lambda_{\bs{v}_{2}\bs{v}_{2'}}\right|^2
(2\pi)^3 
\delta(\epsilon_1\!-\!\epsilon_{1'}\!+\!\omega)
\delta(\bs{k}_1\!-\!\bs{k}_{1'}\!+\!\bs{q})
(2\pi)^3 
\delta(\epsilon_2\!-\!\epsilon_{2'}\!-\!\omega)
\delta(\bs{k}_2\!-\!\bs{k}_{2'}\!-\!\bs{q}),
\end{equation}
where
\begin{equation}
\label{dfs}
\lambda_{\bs{v}_1, \bs{v}_{1'}} = \frac{1}{2}
\left(
1\!+\!\frac{\bs{v}_1\!\cdot\!\bs{v}_{1'}}{v_g^2}
\right)
=
\frac{1}{2}
\left(
1\!+\!\lambda\lambda'\frac{\bs{k}_1\!\cdot\!\bs{k}_{1'}}{k_1k_{1'}}
\right)\!.
\end{equation}
The vertices $\lambda_{\bs{v},\bs{v}'}$ are known
as the ``Dirac factors''. They indicate the asymmetry of
quasi-particle scattering in graphene \cite{kats}.

The integrated collision integrals $\bs{\cal I}_i^{ee}$ are obtained by
multiplying the collision integral ${\rm St}_{ee}$ by
\[
\bs{v}\phi_i, \qquad
\phi_1=1, \qquad
\phi_2=\lambda,
\]
following by integration. The resulting integrated collision integrals
are given by (again, the degeneracy factors are written down
explicitly)
\begin{subequations}
\label{ici}
\begin{equation}
\label{ici0}
\bs{\cal I}_i^{ee}\left[\delta f\right]
=
N^2\sum_{1,1',2,2'} \!\!\bs{v}_2 \phi_{i,2}
W_{12,1'2'} f_{1}^{(0)}f_{2}^{(0)}\left[1\!-\!f_{1'}^{(0)}\right]
  \left[1\!-\!f_{2'}^{(0)}\right]\Big[h_{1'}\!+\!h_{2'}\!-\!h_1\!-\!h_2\Big].
\end{equation}
Due to the time-reversal symmetry of the theory, the transition
probability is symmetric under the interchange of the ``in'' and
``out'' variables and hence one may re-write Eq.~(\ref{ici0}) as
\begin{equation}
\label{ici01}
\bs{\cal I}_i^{ee}\left[\delta f\right]
=
\frac{N^2}{2} \sum_{1,1',2,2'}\!\!
(\bs{v}_2 \phi_{i,2}-\bs{v}_{2'} \phi_{i,2'})
W_{12,1'2'} f_{1}^{(0)}f_{2}^{(0)}\left[1\!-\!f_{1'}^{(0)}\right]
\left[1\!-\!f_{2'}^{(0)}\right]\Big[h_{1'}\!+\!h_{2'}\!-\!h_1\!-\!h_2\Big].
\end{equation}
Using the Golden Rule expression for the transition probability (\ref{w}), I
re-write the collision integral as
\begin{eqnarray}
&&
\bs{\cal I}_i^{ee}\left[\delta f\right]
=
\frac{N^2}{2}
\int\frac{d^2q}{(2\pi)^2}\frac{d\omega}{2\pi}
|U(\omega,\bs{q})|^2
\sum_{1,1'}
  (2\pi)^3 \left|\lambda_{\bs{v}_{1}\bs{v}_{1'}}\right|^2
  \delta(\epsilon_1-\epsilon_{1'}+\omega)
  \delta(\bs{k}_1-\bs{k}_{1'}+\bs{q})
  f_{1}^{(0)} \left[1\!-\!f_{1'}^{(0)}\right]
\nonumber\\
&&
\nonumber\\
&&
\qquad\qquad\qquad\qquad
\times
\sum_{2,2'}
  (2\pi)^3 \left|\lambda_{\bs{v}_{2}\bs{v}_{2'}}\right|^2
  \left(\bs{v}_2\phi_{i,2}\!-\!\bs{v}_{2'}\phi_{i,2'}\right)
  \delta(\epsilon_2-\epsilon_{2'}-\omega)
  \delta(\bs{k}_2-\bs{k}_{2'}-\bs{q})
  f_{2}^{(0)} \left[1\!-\!f_{2'}^{(0)}\right]
\nonumber\\
&&
\nonumber\\
&&
\qquad\qquad\qquad\qquad
\times
\Big[h_{1'}\!+\!h_{2'}\!-\!h_1\!-\!h_2\Big].
\label{ici1}
\end{eqnarray}
\end{subequations}

Substituting the nonequilibrium distribution function (\ref{hs0}), one
finds after a straightforward but tedious calculation (the integral
$\bs{\cal I}_3^{ee}=0$ is introduced for consistency of the notation)
\begin{eqnarray}
\label{li1r}
\begin{pmatrix}
  \bs{\cal I}_1^{ee} \cr
  \bs{\cal I}_2^{ee} \cr
  \bs{\cal I}_3^{ee}
\end{pmatrix}
=
- \frac{1}{2}v_gT\frac{\partial n}{\partial\mu}
\begin{pmatrix}
  \tau^{-1}_{11} & \tau^{-1}_{12} & 0 \cr
  \tau^{-1}_{12} & \tau^{-1}_{22} & 0 \cr
  0 & 0 & 0
\end{pmatrix}
\begin{pmatrix}
  \bs{h}^{(1)} \cr
  \bs{h}^{(2)} \cr
  \bs{h}^{(3)}
\end{pmatrix}
=
-
\frac{\partial n}{\partial\mu}
\begin{pmatrix}
  \tau^{-1}_{11} & \tau^{-1}_{12} & 0 \cr
  \tau^{-1}_{12} & \tau^{-1}_{22} & 0 \cr
  0 & 0 & 0
\end{pmatrix}
\widehat{M}_h^{-1}
\begin{pmatrix}
  \delta\bs{j} \cr
  \delta\bs{j}_I \cr
  \delta\bs{j}_E/T
\end{pmatrix}.
\end{eqnarray}
In the case ${\mu_\pm=\mu}$ considered in this paper, the inverse
times $\tau^{-1}_{ij}$ are given by the following integrals:
\begin{equation}
\label{taus2}
\frac{1}{\tau_{ij}}
=
\pi^2 \alpha_g^2NT \left[\frac{NT}{v_g^2\partial n/\partial\mu}\right]
\int\frac{d^2Q}{(2\pi)^2}\frac{dW}{2\pi}
\frac{|\widetilde{U}|^2}{\sinh^2W}
\Big[
Y_{00}(W,\bs{Q})Y_{ij}(W,\bs{Q})\!-\! Y_{0j}(W,\bs{Q})Y_{0i}(W,\bs{Q})  
\Big].
\end{equation}
Hereafter I use dimensionless variables (the dimensionless frequency
$W$ should not be confused with the enthalpy)
\begin{equation}
\label{dv}
\bs{Q} = \frac{v_g\bs{q}}{2T},
\qquad
W = \frac{\omega}{2T},
\qquad
\Omega = \frac{W}{Q},
\qquad
x=\frac{\mu}{T}.
\end{equation}
The Coulomb interaction has the form
\begin{equation}
\label{coulomb}
U(\omega,\bs{q})= \frac{2\pi e^2}{q} \widetilde{U} = \frac{2\pi\alpha_g v_g}{q}\widetilde{U}, 
\qquad
\alpha_g=\frac{e^2}{v_g\varepsilon},
\end{equation}
where $\varepsilon$ is the effective dielectric constant describing
the electrostatic environment and $\widetilde{U}$ accounts for
screening effects.

The auxiliary functions $Y$ in Eq.~(\ref{taus2}) are given by
\begin{subequations}
  \label{ys_c}
  \begin{equation}
  \label{y00i}
  Y_{00}(\omega,\bs{q}) = \frac{1}{4\pi}
  \left[
    \frac{\theta(|\Omega|\leqslant1)}{\sqrt{1\!-\!\Omega^2}}\,
    {\cal Z}_0^>[I_1]
    +
    \frac{\theta(|\Omega|\geqslant1)}{\sqrt{\Omega^2\!-\!1}}\,
    {\cal Z}_0^<[I_1]
    \right]\!,
\end{equation}
\begin{equation}
  \label{y01i}
  Y_{01}(\omega,\bs{q}) = -\frac{1}{2\pi}
  \left[
    \theta(|\Omega|\leqslant1)\sqrt{1\!-\!\Omega^2}\,
    {\cal Z}_2^>[I]
    +
    \theta(|\Omega|\geqslant1)\sqrt{\Omega^2\!-\!1}\,
    {\cal Z}_2^<[I]
    \right]\!,
\end{equation}
\begin{equation}
  \label{y02i}
  Y_{02}(\omega,\bs{q}) = \frac{1}{2\pi}
  \left[
    \theta(|\Omega|\leqslant1)\sqrt{1\!-\!\Omega^2}\,
    {\cal Z}_2^>[I_1]
    +
    \theta(|\Omega|\geqslant1)\frac{|\Omega|}{\sqrt{\Omega^2\!-\!1}}\,{\cal Z}^<_3[I_1]
    \right],
\end{equation}
\begin{eqnarray}
\label{y11i}
Y_{11}(\omega,\bs{q}) = \frac{1}{\pi}
\left[
     \theta(|\Omega|\leqslant1)\sqrt{1\!-\!\Omega^2}\,
    {\cal Z}_1^>[I_1]
    +
    \theta(|\Omega|\geqslant1)\sqrt{\Omega^2\!-\!1}\,
    {\cal Z}_1^<[I_1]
    \right],
\end{eqnarray}
\begin{equation}
  \label{y12i}
  Y_{12}(\omega,\bs{q}) =- \frac{1}{\pi}
    \theta(|\Omega|\leqslant1)\sqrt{1\!-\!\Omega^2}\,
    {\cal Z}_1^>[I],
\end{equation}
\begin{equation}
  \label{y22i}
  Y_{22}(\omega,\bs{q}) = \frac{1}{\pi}
  \left[
    \theta(|\Omega|\leqslant1)\sqrt{1\!-\!\Omega^2}\,
    {\cal Z}_1^>[I_1]
    +
    \frac{\theta(|\Omega|\geqslant1)}{\sqrt{\Omega^2\!-\!1}}\,{\cal Z}^<_3[I_1]
\right],
\end{equation}
\end{subequations}
where
\begin{subequations}
\label{Zints}
\begin{equation}
\label{z0}
{\cal Z}^>_0[I] = \int\limits_1^\infty dz\sqrt{z^2\!-\!1} \, I(z),
\qquad
{\cal Z}^<_0[I] = \int\limits_0^1 dz\sqrt{1\!-\!z^2} \, I(z),
\end{equation}
\begin{equation}
\label{z1}
{\cal Z}^>_1[I] = \int\limits_1^\infty dz \frac{\sqrt{z^2\!-\!1}}{z^2\!-\!\Omega^2} \, I(z),
\qquad
{\cal Z}^<_1[I] = \int\limits_0^1 dz \frac{\sqrt{1\!-\!z^2}}{\Omega^2\!-\!z^2}\, I(z),
\end{equation}
\begin{equation}
\label{z2}
{\cal Z}^>_2[I] = \int\limits_1^\infty dz \frac{z\sqrt{z^2\!-\!1}}{z^2\!-\!\Omega^2} \, I(z),
\qquad
{\cal Z}^<_2[I] = \int\limits_0^1 dz \frac{z\sqrt{1\!-\!z^2}}{\Omega^2\!-\!z^2}\, I(z),
\end{equation}
\begin{equation}
\label{z3}
{\cal Z}^>_3[I] = \int\limits_1^\infty dz \frac{\left(z^2\!-\!1\right)^{3/2}}{z^2\!-\!\Omega^2} \, I(z),
\qquad
{\cal Z}^<_3[I] = \int\limits_0^1 dz \frac{\left(1\!-\!z^2\right)^{3/2}}{\Omega^2\!-\!z^2}\, I(z),
\end{equation}
\end{subequations}
and
\begin{subequations}
\label{izs}
\begin{equation}
\label{i_10}
I_1(z) = \tanh\frac{zQ+W+x}{2}+\tanh\frac{zQ+W-x}{2}-\tanh\frac{zQ-W+x}{2}-\tanh\frac{zQ-W-x}{2},
\end{equation}
\begin{equation}
\label{iz0}
I(z) = \tanh\frac{zQ+W+x}{2}-\tanh\frac{zQ+W-x}{2}-\tanh\frac{zQ-W+x}{2}+\tanh\frac{zQ-W-x}{2}.
\end{equation}
\end{subequations}

\subsubsection{Contribution of the collision integral due to disorder}

In this paper, I am using the simplest $\tau$-approximation for the
collision integral due to disorder scattering, see Eq.~(\ref{ke}). The
corresponding integrated collision integral is given by
\begin{subequations}
\label{intcid}
\begin{equation}
\bs{\cal I}_i^{\rm dis} = -N\sum_\lambda\int\frac{d^2k}{(2\pi)^2}
\bs{v}_{\lambda\bs{k}}\phi_i
\frac{f^{(0)}_{\lambda\bs{k}}\!+\!\delta f_{\lambda\bs{k}}
-\left\langle f^{(0)}_{\lambda\bs{k}}\!+\!\delta f_{\lambda\bs{k}} \right\rangle_\varphi}{\tau_{\rm dis}},
\end{equation}
such that
\begin{equation}
\bs{\cal I}_1^{\rm dis} = -\frac{\bs{j}}{\tau_{\rm dis}},
\quad
\bs{\cal I}_2^{\rm dis} = -\frac{\bs{j}_I}{\tau_{\rm dis}},
\quad
\bs{\cal I}_3^{\rm dis} = -\frac{\bs{j}_E}{T\tau_{\rm dis}}.
\end{equation}
\end{subequations}
Here I have used the form (\ref{df})-(\ref{hs}) of the non-equilibrium correction
to the distribution function.

\subsection{Collision integral in the tensor equations}
\label{citeqs}

\subsubsection{Collision integral due to electron-electron interaction}

Integrating the collision integral with the factors 
$v^\alpha v^\beta$, $\lambda v^\alpha v^\beta$, and $\epsilon v^\alpha
v^\beta/T$ yields the following tensor quantities, see Eq.~(\ref{teneqs})
\begin{subequations}
\label{cits}
\begin{equation}
\label{i1ab}
{\cal I}^{\alpha\beta}_1= N\sum_\lambda\int\!\frac{d^2k}{(2\pi)^2}
{v}^\alpha_{\lambda\bs{k}} {v}^\beta_{\lambda\bs{k}}{\rm St}_{ee},
\end{equation}
\begin{equation}
\label{i2ab}
{\cal I}^{\alpha\beta}_2= N\sum_\lambda\lambda\int\!\frac{d^2k}{(2\pi)^2}
{v}^\alpha_{\lambda\bs{k}} {v}^\beta_{\lambda\bs{k}}{\rm St}_{ee},
\end{equation}
\begin{equation}
\label{i3ab}
{\cal I}^{\alpha\beta}_3= \frac{N}{T}\sum_\lambda\int\!\frac{d^2k}{(2\pi)^2}
\epsilon_{\lambda\bs{k}} {v}^\alpha_{\lambda\bs{k}} {v}^\beta_{\lambda\bs{k}}{\rm St}_{ee}.
\end{equation}
\end{subequations}
Combining the multiplication factors using the ``mode'' notations, see
Eq.~(\ref{hs0}),
\[
v^\alpha v^\beta \phi_i, \qquad
\phi_1=1, \qquad
\phi_2=\lambda, \qquad
\phi_3=\epsilon/T,
\]
and following the same steps as in \ref{ciceqs}, I obtain the
expression
\begin{eqnarray}
\label{icit}
&&
{\cal I}_i^{\alpha\beta}\left[\delta f\right]
= \sum_{j=1}^3 {h}^{(j)}_{\gamma\delta} \times
\frac{N^2}{2}
\int\frac{d^2q}{(2\pi)^2}\frac{d\omega}{2\pi}
|U(\omega,\bs{q})|^2
\\
&&
\nonumber\\
&&
\qquad
\times
\Bigg[
\frac{(2\pi)^3}{v_g}\sum_{1,1'}
\left({v}^\gamma_{1'}{v}^\delta_{1'}\phi_{j,1'}\!-\!{v}^\gamma_{1'}{v}^\delta_{1}\phi_{j,1}\right)
  \left|\lambda_{\bs{v}_{1}\bs{v}_{1'}}\right|^2
  \delta(\epsilon_1\!-\!\epsilon_{1'}\!+\!\omega)
  \delta(\bs{k}_1\!-\!\bs{k}_{1'}\!+\!\bs{q})
  f_{1}^{(0)} \left[1\!-\!f_{1'}^{(0)}\right]
\nonumber\\
&&
\nonumber\\
&&
\qquad\qquad\qquad
\times
\frac{(2\pi)^3}{v_g}\sum_{2,2'}
  \left({v}^\alpha_{2'}{v}^\beta_{2'}\phi_{i,2'}\!-\!{v}^\alpha_{2}{v}^\beta_{2}\phi_{i,2}\right)
  \left|\lambda_{\bs{v}_{2}\bs{v}_{2'}}\right|^2
  \delta(\epsilon_2\!-\!\epsilon_{2'}\!-\!\omega)
  \delta(\bs{k}_2\!-\!\bs{k}_{2'}\!-\!\bs{q})
  f_{2}^{(0)} \left[1\!-\!f_{2'}^{(0)}\right]
\nonumber\\
&&
\nonumber\\
&&
\qquad\quad
+
(2\pi)^3v_g\sum_{1,1'}
  \left|\lambda_{\bs{v}_{1}\bs{v}_{1'}}\right|^2
  \delta(\epsilon_1\!-\!\epsilon_{1'}\!+\!\omega)
  \delta(\bs{k}_1\!-\!\bs{k}_{1'}\!+\!\bs{q})
  f_{1}^{(0)} \left[1\!-\!f_{1'}^{(0)}\right]
\nonumber\\
&&
\nonumber\\
&&
\qquad\qquad\qquad
\times
\frac{(2\pi)^3}{v_g^3}\sum_{2,2'}
\!\left({v}^\alpha_{2'}{v}^\beta_{2'}\phi_{i,2'}\!-\!{v}^\alpha_{2}{v}^\beta_{2}\phi_{i,2}\right)\!
\left({v}^\gamma_{2'}{v}^\delta_{2'}\phi_{j,2'}\!-\!{v}^\gamma_{2'}{v}^\delta_{2}\phi_{j,1}\right)
\nonumber\\
&&
\nonumber\\
&&
\qquad\qquad\qquad\qquad\qquad\qquad
\times
  \left|\lambda_{\bs{v}_{2}\bs{v}_{2'}}\right|^2
  \delta(\epsilon_2\!-\!\epsilon_{2'}\!-\!\omega)
  \delta(\bs{k}_2\!-\!\bs{k}_{2'}\!-\!\bs{q})
  f_{2}^{(0)} \left[1\!-\!f_{2'}^{(0)}\right]\!
  \Bigg]\!.
\nonumber
\end{eqnarray}

Evaluating the integrals (for ${\mu_\pm=\mu}$ and ${\bs{u}=0}$) I find
a vector in the ``mode space'' that can be written in the form similar
to Eq.~(\ref{li1r}) as
\begin{equation}
  \label{iab0_1}
\begin{pmatrix}
{\cal I}^{\alpha\beta}_1\left[\delta f\right] \cr
{\cal I}^{\alpha\beta}_2\left[\delta f\right] \cr
{\cal I}^{\alpha\beta}_3\left[\delta f\right]
\end{pmatrix}
=
- \frac{1}{4}v_g^2T\frac{\partial n}{\partial\mu}
\begin{pmatrix}
\tilde\tau_{11}^{-1} & \tilde\tau_{12}^{-1} & \tilde\tau_{13}^{-1} \cr
\tilde\tau_{12}^{-1} & \tilde\tau_{22}^{-1} & \tilde\tau_{23}^{-1} \cr
\tilde\tau_{13}^{-1} & \tilde\tau_{23}^{-1} & \tilde\tau_{33}^{-1}
\end{pmatrix}
\begin{pmatrix}
h^{(1)}_{\alpha\beta} \cr
h^{(2)}_{\alpha\beta} \cr
h^{(3)}_{\alpha\beta}
\end{pmatrix}
=
-\frac{\partial n}{\partial\mu}
\begin{pmatrix}
\tilde\tau_{11}^{-1} & \tilde\tau_{12}^{-1} & \tilde\tau_{13}^{-1}\cr
\tilde\tau_{12}^{-1} & \tilde\tau_{22}^{-1} & \tilde\tau_{23}^{-1} \cr
\tilde\tau_{13}^{-1} & \tilde\tau_{23}^{-1} & \tilde\tau_{33}^{-1}
\end{pmatrix}
\widehat{M}_h^{-1}
\begin{pmatrix}
\delta\Pi^{\alpha\beta} \cr
\delta\Pi^{\alpha\beta}_I \cr
v_g^2 \delta\Pi^{\alpha\beta}_E /T
\end{pmatrix},
\end{equation}
with
\begin{equation}
\label{ttauij}
\frac{1}{\tilde\tau_{ij}} =  
(2\pi)^2 \alpha_g^2NT \left[\frac{NT}{v_g^2\partial n/\partial\mu}\right]
\int\frac{d^2Q}{(2\pi)^2}\frac{dW}{2\pi}
\frac{|\widetilde{U}|^2}{\sinh^2W}
\left[{Y}_{00}\widetilde{Y}_{ij}- \widetilde{Y}_{0j} \widetilde{Y}_{0i}\right]\!.
\end{equation}
Here I have introduced auxiliary functions
\begin{subequations}
\label{tys}
\begin{equation}
  \label{ty_01}
\widetilde{Y}_{01}(\omega,\bs{q})
= -\frac{1}{\pi}\!
  \left[
    \theta(|\Omega|\leqslant1)\Omega\sqrt{1\!-\!\Omega^2}\,
    {\cal Z}^>_5[I]
    +
    \theta(|\Omega|\geqslant1)\Omega\sqrt{\Omega^2\!-\!1}\,
    {\cal Z}^<_5[I]
    \right],
\end{equation}
\begin{equation}
\label{ty_02}
\widetilde{Y}_{02}(\omega,\bs{q})
= 
\frac{\theta(|\Omega|\leqslant1)}{\pi}\Omega\sqrt{1\!-\!\Omega^2}\,
{\cal Z}^>_5[I_1]
-
\frac{\theta(|\Omega|\geqslant1)}{2\pi}\, {\rm sign}(\Omega) \sqrt{\Omega^2\!-\!1}\,
\widetilde{\cal Z}^<_4[I_1]
+
\frac{\theta(|\Omega|\geqslant1)}{4\pi}
\frac{{\rm sign}(\Omega)}{\sqrt{\Omega^2\!-\!1}}\,{\cal Z}^<_0[I_1],
\end{equation}
\begin{equation}
\label{ty_03}
\widetilde{Y}_{03}(\omega,\bs{q})
=
Q\Omega \,Y_{00}(\omega,\bs{q})
+
\frac{1}{2\pi}\,Q\Omega
\left[
\theta(|\Omega|\leqslant1)\sqrt{1\!-\!\Omega^2}
\,{\cal Z}^>_3[I_1]
-
\theta(|\Omega|\geqslant1)\sqrt{\Omega^2\!-\!1}
\,{\cal Z}^<_3[I_1]
\right],
\end{equation}
\begin{equation}
\label{ty_11}
\widetilde{Y}_{11}(\omega,\bs{q})
= \frac{1}{\pi}\!
\left[
\theta(|\Omega|\leqslant1)\sqrt{1\!-\!\Omega^2}\,{\cal Z}^>_4[I_1]
+
\theta(|\Omega|\geqslant1)\sqrt{\Omega^2\!-\!1}\,{\cal Z}^<_4[I_1]
\right]\!,
\end{equation}
\begin{equation}
\label{ty_12}
\widetilde{Y}_{12}(\omega,\bs{q})
= -\frac{1}{\pi}
\theta(|\Omega|\leqslant1)\sqrt{1\!-\!\Omega^2}
\,{\cal Z}^>_4[I],
\end{equation}
\begin{equation}
\label{ty_13}
\widetilde{Y}_{13}(\omega,\bs{q})
= -\frac{Q}{\pi}
\left[
\theta(|\Omega|\leqslant1)\sqrt{1\!-\!\Omega^2}\,{\cal Z}^>_5[I]
+
\theta(|\Omega|\geqslant1)\sqrt{\Omega^2\!-\!1}\,{\cal Z}^<_5[I]
\right]\!,
\end{equation}
\begin{eqnarray}
\label{ty_22}
\widetilde{Y}_{22}(\omega,\bs{q})
= \frac{1}{\pi}\left[
\theta(|\Omega|\leqslant1)\sqrt{1\!-\!\Omega^2}
\,{\cal Z}^>_4[I_1]
-
\theta(|\Omega|\geqslant1)\sqrt{\Omega^2\!-\!1}
\,{\cal Z}^<_4[I_1]
+ \frac{1}{4}
\frac{\theta(|\Omega|\geqslant1)}{\sqrt{\Omega^2\!-\!1}}
\,{\cal Z}^<_0[I_1]
\right],
\end{eqnarray}
\begin{equation}
\label{ty_23}
\widetilde{Y}_{23}(\omega,\bs{q})
= \frac{Q}{\pi}\left[
\theta(|\Omega|\leqslant1)\sqrt{1\!-\!\Omega^2}
\,{\cal Z}^>_5[I_1]
-
\theta(|\Omega|\geqslant1)|\Omega|\sqrt{\Omega^2\!-\!1}
\,{\cal Z}^<_4[I_1]
+ \frac{|\Omega|}{4}
\frac{\theta(|\Omega|\geqslant1)}{\sqrt{\Omega^2\!-\!1}}
\,{\cal Z}^<_0[I_1]
\right]\!,
\end{equation}
\begin{equation}
\label{ty_33}
\widetilde{Y}_{33}(\omega,\bs{q})
=
\frac{Q^2}{\pi}\left[
\theta(|\Omega|\leqslant1)\sqrt{1\!-\!\Omega^2}
\,{\cal Z}^>_3[I_1]
-
\theta(|\Omega|\geqslant1)\sqrt{\Omega^2\!-\!1}
\,{\cal Z}^<_3[I_1]\right]
+Q^2\Omega^2 Y_{00}(\omega,\bs{q}),
\end{equation}
\end{subequations}
that involve the integrals (\ref{Zints}) that need to be complemented by
\begin{subequations}
\label{Zints2}
\begin{equation}
\label{z4}
{\cal Z}^>_4[I] = \int\limits_1^\infty dz
\frac{\left(z^2\!-\!1\right)^{3/2}}{\left(z^2\!-\!\Omega^2\right)^2} \, I(z),
\quad
{\cal Z}^<_4[I] = \int\limits_0^1 dz
\frac{\left(1\!-\!z^2\right)^{3/2}}{\left(\Omega^2\!-\!z^2\right)^2}\, I(z),
\quad
\widetilde{\cal Z}^<_4[I] = \int\limits_0^1 dz
\frac{\left(1\!-\!z^2\right)^{3/2}}{\left(\Omega^2\!-\!z^2\right)^2}
\left(\Omega^2\!+\!z^2\right)\, I(z),
\end{equation}
\begin{equation}
\label{z5}
{\cal Z}^>_5[I] = \int\limits_1^\infty dz
\frac{z\left(z^2\!-\!1\right)^{3/2}}{\left(z^2\!-\!\Omega^2\right)^2} \, I(z),
\qquad
{\cal Z}^<_5[I] = \int\limits_0^1 dz
\frac{z\left(1\!-\!z^2\right)^{3/2}}{\left(\Omega^2\!-\!z^2\right)^2}\, I(z).
\end{equation}
\end{subequations}

\subsubsection{Contribution of the collision integral due to disorder}

Using the simplest $\tau$-approximation for the collision integral due
to disorder scattering, see Eq.~(\ref{ke}), I find the integrated
collision integral in the tensor equations (\ref{teneqs}) as
\begin{subequations}
\label{inttid}
\begin{equation}
{\cal I}_i^{\alpha\beta;\,\rm dis} = -N\sum_\lambda\int\frac{d^2k}{(2\pi)^2}
{v}_{\lambda\bs{k}}^\alpha{v}_{\lambda\bs{k}}^\beta\phi_i
\frac{f^{(0)}_{\lambda\bs{k}}\!+\!\delta f_{\lambda\bs{k}}
-\left\langle f^{(0)}_{\lambda\bs{k}}\!+\!\delta f_{\lambda\bs{k}} \right\rangle_\varphi}{\tau_{\rm dis}},
\end{equation}
such that
\begin{equation}
{\cal I}_1^{\alpha\beta;\,\rm dis} = -\frac{\delta\Pi^{\alpha\beta}}{\tau_{\rm dis}},
\quad
{\cal I}_2^{\alpha\beta;\,\rm dis} = -\frac{\delta\Pi^{\alpha\beta}_I}{\tau_{\rm dis}},
\quad
{\cal I}_3^{\alpha\beta;\,\rm dis} = -\frac{v_g^2\delta\Pi^{\alpha\beta}_E}{T\tau_{\rm dis}}.
\end{equation}
\end{subequations}
Here I have used the form (\ref{df})-(\ref{hs}) of the non-equilibrium correction
to the distribution function and evaluated the collision integral in
the co-moving frame, ${\bs{u}\rightarrow0}$, where
\[
\left\langle f^{(0)}_{\lambda\bs{k}}(\bs{u}\!=\!0)+\delta f_{\lambda\bs{k}}(\bs{u}\!=\!0) \right\rangle_\varphi
=
f^{(0)}_{\lambda\bs{k}}(\bs{u}\!=\!0).
\]

\section{Integrated Lorenz terms}

\subsection{Contribution to the current equations}
\label{kvecs}

To evaluate the contribution of the Lorentz force to the equations for
quasiparticle currents, I need to calculate the quantities $\bs{\cal
  K}$ and $\bs{\cal K}_I$ defined in Eqs.~(\ref{lj04}) and
(\ref{ljI04}). These expressions should be evaluated with the
non-equilibrium correction to the distribution function
(\ref{djs}).

In local equilibrium, both vectors are proportional to the
hydrodynamic velocity. In the limit ${\bs{u}\rightarrow0}$, relevant
for the linear response derivation of the dissipative corrections to
the ideal hydrodynamics, I find
\begin{equation}
\label{ks}
\bs{\cal K}^{(0)}={\cal T}N\sum_\lambda \int\!\frac{d^2k}{(2\pi)^2}
  \frac{\bs{k}}{k^2} f^{(0)}_{\lambda\bs{k}}
\rightarrow \frac{\cal T}{2}\frac{\partial n}{\partial\mu} \bs{u},
\qquad
\bs{\cal K}_I^{(0)}={\cal T}N\sum_\lambda \lambda\!\int\!\frac{d^2k}{(2\pi)^2}
  \frac{\bs{k}}{k^2} f^{(0)}_{\lambda\bs{k}}
\rightarrow \frac{\mu}{2}\frac{\partial n}{\partial\mu} \bs{u}.
\end{equation}
Substituting the non-equilibrium distribution correction
(\ref{df})-(\ref{hs}), I find the correction to the quantity
$\bs{\cal{K}}$
\begin{eqnarray*}
&&
\delta\bs{\cal K} = {\cal T}N\sum_\lambda \!\int\!\frac{d^2k}{(2\pi)^2}
\frac{\bs{k}}{k^2}
\!\left(-T\frac{\partial f_{\lambda\bs{k}}^{(0)}}{\partial\epsilon_{\lambda\bs{k}}}\right)
\!\left(\frac{\bs{v}_{\lambda\bs{k}}}{v_g}\!\left[\bs{h}^{(1)}\!+\!\lambda\bs{h}^{(2)}
\!+\!\lambda v_g\frac{k}{T}\bs{h}^{(3)}\right]\right)
\\
&&
\\
&&
\qquad\qquad\qquad\qquad
=
{\cal T}N\sum_\lambda \lambda
\!\int\!\frac{d^2k}{(2\pi)^2}
\frac{\bs{k}}{k^2}
\!\left(-T\frac{\partial f_{\lambda\bs{k}}^{(0)}}{\partial\epsilon_{\lambda\bs{k}}}\right)
\!\left(\frac{\bs{k}}{k}\!\left[\bs{h}^{(1)}\!+\!\lambda\bs{h}^{(2)}
\!+\!\lambda v_g\frac{k}{T}\bs{h}^{(3)}\right]\right).
\end{eqnarray*}
Evaluating the angular integral as usual, 
\[
\int\!\!\frac{d^2k}{(2\pi)^2}
\frac{k^\alpha k^\beta}{k^3}
= \frac{1}{2}\delta^{\alpha\beta}\!\int\limits_0^\infty\!\frac{dk}{2\pi},
\]
I find
\begin{subequations}
\label{lks}
\begin{equation}
\label{k}
\delta\bs{\cal K} = \frac{1}{2}{\cal T}
\sum_{n=1}^3\bs{h}^{(n)}
N\sum_\lambda\lambda\int\limits_0^\infty\!\frac{dk}{2\pi} \phi_n
\left(-T\frac{\partial f^{(0)}_{\lambda\bs{k}}}{\partial\epsilon_{\lambda\bs{k}}}\right).
\end{equation}
Evaluating the remaining integrals I obtain
\begin{equation}
\delta\bs{\cal K}
=
\frac{v_gT}{2} \frac{\partial n}{\partial\mu}
\left[
\bs{h}^{(1)}
\!\left(
\frac{1}{1\!+\!e^{-\mu_+/T}}+\frac{1}{1\!+\!e^{-\mu_-/T}}-1
\right)
+ \bs{h}^{(2)}
\!\left(
1+\frac{1}{1\!+\!e^{-\mu_+/T}}-\frac{1}{1\!+\!e^{-\mu_-/T}}
\right)
+ \bs{h}^{(3)} \frac{\cal T}{T}
\right].
\end{equation}
For the standard case $\mu_\pm\!=\!\mu$ the result simplifies to
\begin{equation}
\delta\bs{\cal K}(\mu_\pm\!=\!\mu)=\frac{v_gT}{2} \frac{\partial n}{\partial\mu}
\left[
\bs{h}^{(1)} \tanh\frac{\mu}{2T} + \bs{h}^{(2)} + \bs{h}^{(3)} \frac{\cal T}{T}
\right].
\end{equation}
Combining the ``ideal'' and ``dissipative'' contributions, I find the
total vector ${\cal K}$ in the form
\begin{equation}
\label{k1}
\bs{\cal K}=\bs{\cal K}^{(0)}+\delta\bs{\cal K}=
\frac{\cal T}{2}\frac{\partial n}{\partial\mu} \bs{u}+
\frac{v_gT}{2} \frac{\partial n}{\partial\mu}
\left[
\bs{h}^{(1)} \tanh\frac{\mu}{2T} + \bs{h}^{(2)} + \bs{h}^{(3)} \frac{\cal T}{T}
\right].
\end{equation}

The quantity $\delta\bs{\cal K}_I$ is calculated in a similar fashion. The
result is given by
\begin{equation}
\label{ki}
\delta\bs{\cal K}_I =
\frac{v_gT}{2} \frac{\partial n}{\partial\mu}
\left[
\bs{h}^{(1)}
\!\left(
\frac{1}{1\!+\!e^{-\mu_+/T}}-\frac{1}{1\!+\!e^{-\mu_-/T}}+1
\right)
+ \bs{h}^{(2)}
\!\left(
\frac{1}{1\!+\!e^{-\mu_+/T}}+\frac{1}{1\!+\!e^{-\mu_-/T}}-1
\right)
+ \bs{h}^{(3)} \frac{\mu}{T}
\right],
\end{equation}
\begin{equation}
\label{ki1}
\bs{\cal K}_I(\mu_\pm\!=\!\mu)
= 
\bs{\cal K}^{(0)}_I+\delta\bs{\cal K}_I=
\frac{\mu}{2}\frac{\partial n}{\partial\mu} \bs{u}+
\frac{v_gT}{2} \frac{\partial n}{\partial\mu}
\left[
\bs{h}^{(1)}  + \bs{h}^{(2)} \tanh\frac{\mu}{2T} + \bs{h}^{(3)} \frac{\mu}{T}
\right].
\end{equation}
\end{subequations}

The results (\ref{k1}) and (\ref{ki1}) coincide with the results of
Ref.~\cite{hydro0}. Using the notations of \ref{ndflr} these
expressions can be re-written as
\[
\bs{\cal K}=\frac{v_gT}{2} \frac{\partial n}{\partial\mu}
\left[
\bs{h}^{(1)} \tanh\frac{\mu}{2T} + \bs{h}^{(2)} 
+ \left(\bs{h}^{(3)}\!+\!\bs{h}^{(3)}_{\bs{u}}\right) \frac{\cal T}{T}
\right],
\]
\[
\bs{\cal K}_I
=
\frac{v_gT}{2} \frac{\partial n}{\partial\mu}
\left[
\bs{h}^{(1)}  + \bs{h}^{(2)} \tanh\frac{\mu}{2T} + 
\left(\bs{h}^{(3)}\!+\!\bs{h}^{(3)}_{\bs{u}}\right) \frac{\mu}{T}
\right].
\]

\subsection{Contribution to the tensor equations}
\label{ksitens}

Integrating the Lorentz terms in the kinetic equation with the factors
$v^\alpha v^\beta$, $\lambda v^\alpha v^\beta$, and $\epsilon v^\alpha
v^\beta/T$ yields the following tensor quantities, see Eq.~(\ref{teneqs})
\begin{subequations}
\label{ksidef}
\begin{equation}
\label{xi1}
\Xi^{i\beta} =
{\cal T}N\!\sum_\lambda\!\int\!\frac{d^2k}{(2\pi)^2} f_{\lambda\bs{k}}
\frac{{v}^i_{\lambda\bs{k}}{v}^\beta_{\lambda\bs{k}}}{\epsilon_{\lambda\bs{k}}}
=
v_g{\cal T}N\!\sum_\lambda\!\lambda\int\!\frac{d^2k}{(2\pi)^2} f_{\lambda\bs{k}}
\frac{k^ik^\beta}{k^3}
=
\frac{v_g{\cal T}}{4}\!\sum_{n=1}^3\!h^{(n)}_{i\beta}
N\!\sum_\lambda\!\lambda\!\int\limits_0^\infty\!\frac{dk}{2\pi} \phi_n
\!\left(\!-T\frac{\partial f^{(0)}_{\lambda\bs{k}}}{\partial\epsilon_{\lambda\bs{k}}}\right)\!,
\end{equation}
\begin{equation}
\label{xi2}
\Xi_I^{i\beta}=
v_g{\cal T}N\sum_\lambda\int\!\frac{d^2k}{(2\pi)^2} f_{\lambda\bs{k}}
\frac{k^ik^\beta}{k^3}
=
\frac{1}{4}v_g{\cal T}\sum_{n=1}^3h^{(n)}_{i\beta}
N\sum_\lambda\int\limits_0^\infty\!\frac{dk}{2\pi} \phi_n
\left(-T\frac{\partial f^{(0)}_{\lambda\bs{k}}}{\partial\epsilon_{\lambda\bs{k}}}\right)\!,
\end{equation}
\begin{equation}
\label{xi3}
\Xi_E^{i\beta}=
v_g{\cal T}N\sum_\lambda\lambda\int\!\frac{d^2k}{(2\pi)^2} f_{\lambda\bs{k}} 
\frac{\epsilon_{\lambda\bs{k}}}{T}
\frac{k^ik^\beta}{k^3}
=
\frac{v_g^2{\cal T}}{4T}\sum_{n=1}^3h^{(n)}_{i\beta}
N\sum_\lambda\int\limits_0^\infty\!\frac{dk}{2\pi} k \phi_n
\left(-T\frac{\partial f^{(0)}_{\lambda\bs{k}}}{\partial\epsilon_{\lambda\bs{k}}}\right)\!.
\end{equation}
\end{subequations}
Substituting the nonequilibrium distribution function (\ref{hs0} and
evaluating the integrals, I find
\begin{subequations}
\label{ksi1}
\begin{equation}
\Xi^{i\beta} = \frac{N}{4}v_g{\cal T}
\left[
h^{(1)}_{i\beta} \frac{T}{2\pi v_g}\tanh\frac{\mu}{2T}
\!+\!
h^{(2)}_{i\beta} \frac{T}{2\pi v_g}
\!+\!
h^{(3)}_{i\beta} \frac{\cal T}{2\pi v_g}
\right]
=
\frac{v_g^2T}{4} \frac{\partial n}{\partial\mu}
\left[
h^{(1)}_{i\beta} \tanh\frac{\mu}{2T}
\!+\!
h^{(2)}_{i\beta}
\!+\!
h^{(3)}_{i\beta} \frac{\cal T}{T}
\right]\!,
\end{equation}
\begin{equation}
\Xi_I^{i\beta} =
\frac{v_g^2T}{4} \frac{\partial n}{\partial\mu}
\left[
h^{(1)}_{i\beta} 
\!+\!
h^{(2)}_{i\beta} \tanh\frac{\mu}{2T}
\!+\!
h^{(3)}_{i\beta} \frac{\mu}{T}
\right]\!,
\end{equation}
\begin{equation}
\Xi_E^{i\beta} =
\frac{v_g^2T}{4} \frac{\partial n}{\partial\mu}
\left[
h^{(1)}_{i\beta} \frac{\cal T}{T}
\!+\!
h^{(2)}_{i\beta} \frac{\mu}{T}
\!+\!
h^{(3)}_{i\beta} \frac{4\pi v_g^2n_0}{NT^2}
\right]
=
\frac{v_g^2T}{4} \frac{\partial n}{\partial\mu}
\left[
h^{(1)}_{i\beta} \frac{\cal T}{T}
\!+\!
h^{(2)}_{i\beta} x
\!+\!
h^{(3)}_{i\beta} 2\tilde{n}
\right]\!.
\end{equation}
\end{subequations}
These results are summarized in the main text in Eq.~(\ref{xivec}).

\bibliographystyle{elsarticle-num}

\bibliography{viscosity_refs}

\end{document}